\documentclass[prd,preprint,preprintnumbers,superscriptaddress,nofootinbib]{revtex4}
\usepackage[a4paper, hdivide={1.91cm,,1.165cm}, vdivide={1.83cm,,3.6cm}]{geometry}

\usepackage{amstext,amssymb}
\usepackage{amsmath}
\usepackage{graphicx}
\usepackage[hyperfootnotes=true]{hyperref}
\usepackage{xspace}
\usepackage{color}
\usepackage{slashed}

\usepackage{xcolor}
\definecolor{dark-green}{rgb}{0.1,0.7,0.3}

\newcommand{\mwl}{M_{W_L}}
\newcommand{\mwr}{M_{W_R}}

\begin{document}
\title{A comparative study of $0\nu\beta\beta$ decay in symmetric and asymmetric left-right model}
\author{Chayan Majumdar$^{1}$,  Sudhanwa Patra$^{2}$, Prativa Pritimita$^{1}$, Supriya Senapati}
\email{\\chayan@phy.iitb.ac.in \\
sudhanwa@iitbhilai.ac.in \\
prativa@iitb.ac.in \\
supriya@phy.iitb.ac.in}
\affiliation{Department of Physics, Indian Institute of Technology Bombay, Powai, Mumbai-400076 \\
$^{2}$Indian Institute of Technology Bhilai, GEC Campus, Sejbahar, Raipur-492015, Chhattisgarh, India}

\begin{abstract}
\vspace*{0.5cm}
We study the new physics contributions to neutrinoless double beta decay ($0\nu\beta\beta$) in a TeV scale left-right model with spontaneous D-parity breaking mechanism 
where the values of the $SU(2)_L$ and $SU(2)_R$ gauge couplings, $g_L$ and $g_R$ are unequal. 
Neutrino mass is generated in the model via gauge extended inverse seesaw mechanism.
We embed the model in a non-supersymmetric $SO(10)$ GUT with a purpose of quantifying the results 
due to the condition $g_{L} \neq g_{R}$. We compare the predicted numerical values of half life of $0\nu\beta\beta$ decay, effective 
Majorana mass parameter and other lepton number violating parameters for three different cases; (i) for manifest left-right symmetric model ($g_L = g_R$), 
(ii) for left-right model with spontaneous D parity breaking ($g_L \neq g_R$), (iii) for Pati-Salam symmetry with D parity breaking ($g_L \neq g_R$). 
We show how different contributions to $0\nu\beta\beta$ decay are suppressed or enhanced depending upon the values 
of the ratio $\frac{g_R}{g_L}$ that are predicted from successful gauge coupling unification.
\end{abstract}

\maketitle

%***************************************************************************************************************
%***************************************************************************************************************
\section{Introduction}
\label{sec:intro}
The immediate question that followed the discovery of neutrino mass and mixing by oscillation 
experiments \cite{Fukuda:2001nk,Ahmad:2002jz,Ahmad:2002ka,Bahcall:2004mz,Abe:2011sj,Adamson:2011qu,Abe:2011fz,Ahn:2012nd,An:2012eh} and still remains unanswered 
is : `Whether neutrinos are Dirac or Majorana particles?' Even more gripping is the question, `What gives them such a tiny mass?', 
since it is believed Higgs mechanism can't be the one responsible. The seesaw mechanism 
\cite{Minkowski:1977sc,Yanagida:1979as,GellMann:1980vs,Mohapatra:1979ia,Cheng:1980qt,Lazarides:1980nt,Magg:1980ut,Schechter:1980gr,Foot:1988aq,Ma:1998dn} 
which is the minimal approach to explain non-zero neutrino mass presumes them as Majorana fermions. 
If neutrinos are Majorana fermions \cite{Majorana:1937vz} they can initiate a very rare process in nature called neutrinoless double beta decay ($0\nu\beta\beta$): 
${}^{A}_{Z} X \to {}^{A}_{Z+2} Y + 2 e^- $, which clearly violates lepton number by two units \cite{Schechter:1981bd}. Therefore this process if observed unambiguously can confirm 
the Majorana nature of neutrinos and total lepton number violation in nature. The detection of this rare phenomena is the main aim of several 
ongoing experiments that are trying to put a bound on the half life of particular nuclei from which limits on the effective Majorana mass can be obtained easily.

At present, KamLAND-Zen experiment gives the bound on half-life as $T^{0\nu}_{1/2} >1.6 \times 10^{26}$ yrs using $^{136}\mbox{Xe}$ 
\cite{Ozaki:2019uyd} while GERDA gives $T^{0\nu}_{1/2} >8.0 \times 10^{25}$ yrs at 90 \% C.L. using $^{76}\mbox{Ge}$ \cite{Agostini:2017hit}. 
Translating these limits into effective mass bound it turns out to be $0.26-0.6$ eV, whereas the 
Planck collaboration puts a tight limit on the sum of light neutrino masses to be $\leq 0.23$ eV at 95\% C.L. \cite{Aghanim:2018eyx}\footnote{However, the 
current cosmological upper limit on the sum of the neutrino masses is a 
bit tighter than Planck collaboration, ranging from 0.19 eV to 0.12 eV. For more detail discussion on this one may refer refs.\cite{Vagnozzi:2017ovm,Giusarma:2018jei,Giusarma:2016phn}.} and as per KATRIN the 
upper bound on lightest neutrino mass, $m_{\beta}< 2$~eV at 95\% C.L. \cite{Tanabashi:2018oca}. Moreover KATRIN is targeted to advance the sensitivity on $m_{\beta} $ down to 0.2 eV (90\% C.L.) in the near future \cite{Aker:2019uuj, Osipowicz:2001sq, Angrik:2005ep, Arenz:2018kma, Kleesiek:2018mel}. 
Which means any positive signal of $0\nu\beta\beta$ decay at the experiments would definitely indicate some new physics contribution to the process. 

One possible way to have new physics contributions to $0\nu\beta\beta$ decay process other than the standard mechanism is to study the process in 
Left-Right Symmetric Model (LRSM) \cite{Mohapatra:1974gc,Pati:1974yy,Senjanovic:1975rk,Senjanovic:1978ev,Mohapatra:1980yp}, which obeys the gauge 
symmetry $SU(2)_L \times SU(2)_R \times U(1)_{B-L} \times SU(3)_C$. The presence of 
right-handed neutrino, doubly charged Higgs scalar, and the possibility of left-right mixing can facilitate new decay channels for the process.
LRSM has already been exhaustively studied \cite{Pritimita:2016fgr,Heeck:2015qra,Garcia-Cely:2015quu,Patra:2015vmp,Patra:2015qny,Deppisch:2017vne,Hati:2018tge,Dev:2018foq,Chauhan:2019fji} 
in order to explain neutrino mass, 
lepton number violation, lepton flavour violation, dark matter and baryon asymmetry of the universe. Even rich collider phenomenology is expected when 
left-right symmetry breaks at TeV scale 
\cite{Tello:2010am,Barry:2013xxa,Dev:2013vxa,Nemevsek:2011hz,Dev:2014iva,Das:2012ii,Bertolini:2014sua,Dhuria:2015cfa,Borah:2013lva,Chakrabortty:2012mh,Deppisch:2015cua, Majumdar:2018eqz, Bambhaniya:2015ipg, Dev:2014xea}. 

However a different scenario arises when the discrete parity symmetry (D-parity) of a left-right symmetric theory breaks at a high scale and the local $SU(2)_R$ symmetry breaks at 
relatively low scale \cite{Chang:1983fu,Chang:1984uy}. This decoupling of D parity breaking and $SU(2)_R$ symmetry breaking introduces a new scale and as an immediate effect, 
the gauge couplings for $SU(2)_L$ and $SU(2)_R$ gauge groups become unequal, i.e. $g_L \neq g_R$. In ref.\cite{Borah:2010zq} 
a TeV scale left-right model with D-parity breaking has been studied and in ref.\cite{Awasthi:2013ff} such a model has been embedded in a non-SUSY $SO(10)$ GUT with 
Pati-Salam symmetry as the highest intermediate breaking step. The same idea has been extended in ref.\cite{Patra:2014goa} 
to study baryon asymmetry of the universe, neutron-antineutron oscillation and proton decay. But the effect of $g_L \neq g_R$ 
has not been emphasized in these above mentioned works while studying $0\nu\beta\beta$ decay. This deviation ($g_L \neq g_R$) brings a noticeable 
difference in the  $0\nu\beta\beta$ decay sector which is the  main essence of this work. We show how different 
contributions to $0\nu\beta\beta$ decay are suppressed or enhanced depending upon the values 
of the ratio $\frac{g_R}{g_L}$ that appears in Feynman amplitudes and $\frac{g_L}{g_R}$ that appears in half-life estimation. 
In order to quantify the results due to unequal $g_L$ and $g_R$ we consider two different symmetry breaking chains from non-SUSY SO(10) GUT; 
one with Pati-Salam symmetry as the highest intermediate step, another without Pati-Salam symmetry \cite{Arbelaez:2013nga}. The importance of Pati-Salam symmetry as the highest 
intermediate step in a $SO(10)$ symmetry breaking chain has already been discussed in ref \cite{Chang:1984uy,Parida:1996td,Nayak:2013dza}.

The rest of the paper is organised as follows. We start our discussion with the basic differences between generic LRSM and 
asymmetric LRSM and present the symmetry breaking steps in section \ref{sec:LRSM}. Next in section \ref{sec:neutrino_mass} we explain how neutrino mass is generated via low scale extended inverse seesaw mechanism in 
the model. We perform the numerical estimation of different $0\nu\beta\beta$ 
contributions in section \ref{sec:numerical_estimation} followed by a comparative study of the process in symmetric and asymmetric left-right case in section \ref{sec:comparative_study}. We summerize our results and conclude in section \ref{sec:conclusion}. In appendix \ref{sec:a1} we discuss the new physics contributions to $0\nu\beta\beta$ decay process that arise in a left-right model with spontaneous D parity breaking. We derive upper limits for different lepton number violating parameters in appendix \ref{sec:lifetime}. We discuss the role of Pati-Salam symmetry in predicting half-life 
of $0\nu\beta\beta$ decay and other LNV parameters and show the gauge coupling unifications in appendix \ref{sec:unification}.

\section{Left-right model with spontaneous D-parity breaking}
\label{sec:LRSM}
In this section we discuss the properties of left-right model with spontaneous D-parity breaking mechanism and state how it  
differs from the manifest left-right model. 
\subsection{Symmetric left-right model}
The symmetric left-right model is based on the gauge group $SU(2)_L \times SU(2)_R \times U(1)_{B-L} \times SU(3)_C$ 
plus a discrete left-right parity symmetry. In these models, the discrete parity and $SU(2)_R$ gauge symmetry break simultaneously and thus the gauge couplings for $SU(2)_L$ and $SU(2)_R$ gauge groups remain equal i.e, $g_{L}=g_{R}$. The fermions and Higgs scalars are related in this model as,

\begin{eqnarray}
&\mbox{Fermions:}& \psi_{L} \equiv \begin{pmatrix}\nu_L \\ \ell_L\end{pmatrix} 
\stackrel{P}{\Longleftrightarrow} \begin{pmatrix}\nu_R \\ \ell_R\end{pmatrix} \equiv \psi_{R}\, ,\quad 
Q_{L} = \begin{pmatrix}u_R \\ d_R\end{pmatrix}
\stackrel{P}{\Longleftrightarrow} Q_{R} =\begin{pmatrix}u_R \\ d_R\end{pmatrix} \, \nonumber \\
&\mbox{Scalars:\,}&
\Delta_{L,R} \equiv \begin{pmatrix} \delta_{L,R}^+/\sqrt{2} & \delta_{L,R}^{++} \\ \delta_{L,R}^0 & -\delta_{L,R}^+/\sqrt{2} \end{pmatrix}\, \quad 
\phi \equiv \begin{pmatrix} \phi_1^0 & \phi_2^+ \\ \phi_1^- & \phi_2^0 \end{pmatrix}\, .
\end{eqnarray}
The symmetry breaking of the left-right symmetric theory down to Standard Model and further to $G_{13} \equiv U(1)_Y \times SU(3)_C$ is achieved by 
the scalars $\Delta_{R,L}$ and $\Phi$ respectively. For a more detailed discussion on spontaneous symmetry breaking of left-right symmetry 
one may refer \cite {Pritimita:2016fgr}.
The vacuum expectation values (vev) of the scalars are given by,
\begin{center}
$\langle \Delta_{L,R} \rangle = 
\begin{pmatrix}
0 & 0 \\
v_{L,R} & 0
\end{pmatrix}$, ~~~
$\langle \Phi \rangle = 
\begin{pmatrix}
k_1 & 0 \\
0 & k_2
\end{pmatrix}$
\end{center}

The invariant Yukawa Lagrangian under the symmetric left-right theory is
\begin{eqnarray}
\mathcal{L}_{\rm Yuk}&=& h_{ij} \overline{\nu}_{L}\, \langle \phi \rangle  N_{jR} 
                  + \tilde{h}_{ij} \overline{\nu}_{L}\, \langle \tilde{\phi} \rangle  N_{jR}
                 +f_{ij} \left[N^T_{i R} C \langle \Delta_R \rangle N_{jR}
                   + \nu^T_{i L} C \langle \Delta_L \rangle \nu_{jL}\right]+\mbox{h.c.} \nonumber \\
&\subset& h_{ij} \overline{\psi}_{iL} \phi \psi_{jR}+ \tilde{h}_{ij} \overline{\psi}_{iL} \tilde{\phi} \psi_{jR}
          +f_{ij} \left[\psi^T_{i} C i\tau_2 \Delta_R \psi_{jR} + R \leftrightarrow L\right]+\mbox{h.c.}
\end{eqnarray}
with $\tilde{\phi} \equiv \tau_2 \phi^{\ast} \tau_2$. The discrete left-right symmetry also results in equal Majorana couplings for left-handed and right-handed neutrinos. 
With these Yukawa terms the neutrino mass formula can be written as, 
\begin{eqnarray}
m_{\nu} = M_L - M_D \frac{1}{M_R} M^T_D = m^{II}_\nu + m^{I}_\nu\, ,
\end{eqnarray}
where $M_D$ is the Dirac neutrino mass matrix, $M_L=f\, v_L =f\,\langle \Delta_L \rangle_{\rm induced}$ 
($M_R=f\, v_R =f\,\langle \Delta_R \rangle$) is the Majorana mass term for left-handed (right-handed) neutrinos.

The seesaw relation in this case is found (from minimization of scalar potential consists of $\Phi, \Delta_{L,R}$) to be,
\begin{equation}
v_L v_R = \gamma k^2\, . 
\end{equation}
where $k = \sqrt{k_1^2 +k_2^2}$ and $\gamma$ represents the function of various scalar coupling parameters in potential.

This means if one assign low mass to $\nu_L$ i.e. around eV scale, then $\nu_R$ has to lie at a very heavy scale, 
say $10^{13}$ to $10^{14}$ GeV which is well beyond the reach of current experiments.
In order to bring down the right-handed scale to TeV, parity and $SU(2)_R$ have to be broken at TeV scale and also the Higgs couplings have to be 
finetuned to order of $\gamma \leq {\cal O}(10^{-10})$.

\subsection{Asymmetric left-right model}
Left-right theory with spontaneous D-parity breaking 
mechanism \cite{Chang:1983fu,Chang:1984uy} is based on the idea that the discrete left-right symmetry (D parity) breaking scale and local $SU(2)_R$ breaking scale are decoupled from each other, i.e, D-parity breaks at an earlier stage as compared to $SU(2)_R$ gauge symmetry, thereby introducing a new scale. It should be noted here that, D-parity should not be confused with the Lorentz parity as latter one acts only on the fermionic content of the theory while D-parity interchanges the parity of the fermion as well as the $SU(2)_L \times SU(2)_R$ Higgs fields. It results in an asymmetric LR model  for which the $SU(2)_L$ and $SU(2)_R$ gauge couplings become unequal i.e, $g_L \neq g_R$. 
This spontaneous breaking of D-parity occurs when singlet scalar $\sigma$ takes vev which is odd under D-parity. 
The asymmetric left-right model then breaks down to SM symmetry with the help of right-handed triplet Higgs scalar $\Delta_R$. Further SM symmetry breaks down 
to $U(1)_{\text{em}}$ theory with the help of scalar bidoublet $\Phi$. The complete symmetry breaking can be sketched as follows,

\begin{centering}
\begin{itemize}
\item $SU(2)_L \times SU(2)_R \times U(1)_{B-L}\times SU(3)_C\times D 
\stackrel{\langle \sigma \rangle}{\longrightarrow} SU(2)_L \times SU(2)_R \times U(1)_{B-L}\times SU(3)_C$ \\
\item $SU(2)_L \times SU(2)_R \times U(1)_{B-L} \times SU(3)_C \stackrel{\langle \Delta_R \rangle} {\longrightarrow} 
SU(2)_L \times U(1)_Y \times SU(3)_C$ \\
\item $SU(2)_L \times U(1)_Y \times SU(3)_C\stackrel{\langle \Phi \rangle}{\longrightarrow}U(1)_{\rm em} \times SU(3)_C$
\end{itemize}
\end{centering}

As $SU(2)_R$ breaking scale and parity breaking scale are different, there is no effect of left-handed scalar
$\Delta_L$ at low energy and the fermion masses can be derived from the Yukawa Lagrangian. However, one can write 
an induced VEV for $\Delta_L$ from the seesaw relation as
\begin{equation}
v_L \approx \frac{\beta k^2v_R}{2M\, M_P}\,
\label{vev_value} 
\end{equation}
where $M_P$ is D-parity breaking scale, $\beta$ is a Higgs coupling constant of $\mathcal{O}(1)$ and $M \simeq M_P$.

This asymmetric left-right gauge theory can also emerge from high scale Pati-Salam theory having gauge group $SU(2)_L 
\times SU(2)_R \times SU(4)_C \times D$, i.e. $ \mathcal{G}_{224D}$. Hence, we cite here two separate models implemented with spontaneous D-parity 
breaking mechanism, both having TeV scale asymmetric left-right model as an intermediate breaking step.
\begin{eqnarray}
&\mbox{I.\,}&\text{SO(10)} \stackrel{M_U}{\longrightarrow}\mathcal{G}_{2213D}\,
       \mathop{\longrightarrow}^{M_P}_{} \mathcal{G}_{2213} \, \stackrel{M_R}{\longrightarrow} G_{\text{SM}} \stackrel{M_Z}{\longrightarrow} G_{13}\, \nonumber \\
&\mbox{II.\,}&\text{SO(10)} \stackrel{M_U}{\longrightarrow}\mathcal{G}_{224D} \stackrel{M_P}{\longrightarrow}  \mathcal{G}_{224}\stackrel{M_C}{\longrightarrow}  \mathcal{G}_{2213} \stackrel{M_R}{\longrightarrow} G_{\text{SM}}\stackrel{M_Z}{\longrightarrow} G_{13}
\end{eqnarray}
Here, $M_U$ represents the unification scale (GUT scale), $M_P, M_C, M_R$ and $M_Z$ correspond to $D-$parity breaking, $SU(4)_C \rightarrow SU(3)_C \times U(1)_{B-L}$ breaking, LR-breaking and SM breaking scale respectively.

\section{Neutrino masses and mixing via extended inverse seesaw mechanism}
\label{sec:neutrino_mass}
In manifest left-right symmetric models where neutrino mass is generated via type-I+type-II seesaw 
mechanism \cite{Chakrabortty:2012mh,Dev:2013vxa,Barry:2013xxa,Das:2012ii,Bertolini:2014sua,Borah:2016iqd,Borah:2015ufa}, 
one has to add either extra symmetry or do structural cancellation in order to align the right handed scale at TeV range. However 
the canonical seesaw contribution can be exactly cancelled out in case of extended type-II seesaw \cite{Patra:2014pga} or 
inverse seesaw mechanism \cite{LalAwasthi:2011aa, Dev:2009aw, Dev:2015pga} and a large value of Dirac neutrino mass matrix, $M_D$ can be obtained. 
These choices of seesaw allow large light-heavy neutrino mixing which facilitate rich phenomenology. However, generic inverse 
seesaw mechanism as proposed in ref. \cite{Awasthi:2013we, Humbert:2015yva} gives negligible 
contribution to neutrinoless double beta decay as the associated sterile neutrino mass matrix  
$\mu_S$ lies in keV range to account for neutrino mass mechanism. Thus one has to extend 
the inverse seesaw mechanism with a large lepton number violating parameter in $N-N$ sector as 
$M_N$ while keeping the same keV value of $\mu_S$ in the $S-S$ sector. Hence, the corresponding seesaw 
mechanism is termed as \textquotedblleft extended inverse seesaw mechanism (EISS) \textquotedblright ~ where the
neutrino mass is governed by the standard inverse seesaw formula. 

Henceforward we consider the model discussed in ref.\cite{Awasthi:2013ff} for our comparative study throughout the paper. 
In this model, gauged inverse seesaw mechanism is implemented by adding one extra fermion singlet $S_i~ (i=1,2,3)$ 
per fermion generation. The extended seesaw mechanism is further gauged at TeV scale for which the VEV of the 
RH-doublet $\langle \chi^0_R \rangle=v_\chi$ 
provides the $N-S$ mixing matrix $M$. 

The asymmetric low-scale Yukawa Lagrangian can be written as,
\begin{eqnarray}
\mathcal{L}_{\rm Yuk} &= & Y^{\ell} \overline{\ell}_L\, N_R\, \Phi 
                       + f\, N^c_R\, N_R \Delta_R 
                       + F\, \overline{N}_R\, S\, \chi_R + S^T \mu_S S +\text{h.c.} 
\end{eqnarray}
which gives rise to the $9\times 9$ neutral lepton mass matrix in the basis of $\begin{pmatrix}
\nu_L & N_R & S
\end{pmatrix}^T$ after electroweak symmetry breaking 
\begin{equation}
\mathcal{M}= \left( \begin{array}{ccc}
              0      & M_D   & 0   \\
              M^T_D  & M_R   & M^T \\
              0      & M     & \mu_S
                      \end{array} \right) \, ,
\label{eqn:numatrix}       
\end{equation}
where $M_D=Y^\ell \langle \Phi\rangle$, $M_R=f\langle \Delta^0_R \rangle$, $M=F\langle\chi_R^0\rangle$. 
The Dirac neutrino mass matrix $M_D$ is determined from the high scale symmetry and fits to charged fermion masses 
at GUT scale using RG evolution equations. In principle the $N-S$ mixing matrix $M$ can assume 
any arbitrary form though we have taken it as diagonal. We have also treated the heavy 
RH Majorana neutrino mass matrix $M_R$ to be diagonal throughout this work. One essential outcome of extended 
inverse seesaw mechanism is that type-I seesaw contribution is exactly canceled out, type-II contribution 
is also damped out and thus inverse seesaw is the only viable contribution to light neutrino masses 
\begin{eqnarray}
m_\nu=\left(\frac{M_D}{M}\right)\,\mu_S\, \left(\frac{M_D}{M}\right)^T\, ,
\end{eqnarray}
The heavy sterile neutrinos and heavy RH neutrinos with their mass matrices can be noted as,
$M_{S}=\mu_S- M M_R^{-1} M^T$,~$M_N =M_R+\cdots$.

As stated earlier, these block diagonal mass matrices $m_\nu$, $M_S$ and $M_N$ can 
further be diagonalized to give physical masses to all neutral leptons by respective unitary 
mixing matrices: $U_\nu$, $U_{S}$ and $U_{N}$ where
\begin{eqnarray}
U^\dagger_\nu\, m_{\nu}\, U^*_{\nu}  &=& m^d_\nu = 
         \text{diag}\left[m_{\nu_1}, m_{\nu_2}, m_{\nu_3}\right]\, , \nonumber \\
U^\dagger_S\, M_{S}\, U^*_{S}  &=& M^d_S = 
         \text{diag}\left[M_{S_1}, M_{S_2}, M_{S_3}\right]\, , \nonumber \\
U^\dagger_N\, M_{N}\, U^*_{N}  &=& M^d_N = 
         \text{diag}\left[M_{N_1}, M_{N_2}, M_{N_3}\right]\,
\label{eq:nudmass}
\end{eqnarray}
The complete mixing matrix \cite{Grimus:2000vj,Mitra:2011qr,Parida:2012sq,Awasthi:2013ff} diagonalizing the resulting $9 \times 9$ neutrino mass matrix turns out to be
\begin{align}
\mathcal{V} & \equiv 
\begin{pmatrix}
\mathcal V^{\nu \nu} & \mathcal V^{\nu S} & \mathcal V^{\nu N} \\
\mathcal V^{S \nu} & \mathcal V^{S S} & \mathcal V^{S N} \\
\mathcal V^{N \nu} & \mathcal V^{N S} & \mathcal V^{N N} 
\end{pmatrix}  \nonumber \\
& = 
\begin{pmatrix}
\left(1-\frac{1}{2}XX^\dagger \right) U_\nu  & 
\left(X-\frac{1}{2}ZY^\dagger \right) U_{S} & 
Z\,U_{N}     \\
-X^\dagger\, U_\nu   &
\left(1-\frac{1}{2} \{X^\dagger X + YY^\dagger \}\right) U_{S} &
\left(Y-\frac{1}{2} X^\dagger Z\right) U_{N}   \\
y^*\, X^\dagger\, U_{\nu} &
-Y^\dagger\, U_{S} &
\left(1-\frac{1}{2}Y^\dagger Y\right)\, U_{N}
\label{eqn:Vmix-extended}
\end{pmatrix}
\end{align}
where the symbols are expressed in terms of model parameters as $X = M_D\,M^{-1}$, $Y=M\, 
M^{-1}_N$, $Z=M_D\,M^{-1}_N$, and $y=M^{-1}\,\mu_S$. With this mixing matrix, one can write 
the relevant charged current interactions of leptons valid at TeV scale asymmetric LR 
gauge theory (with $g_{L} \neq g_{R}$) in the flavor basis as
{\small 
\begin{eqnarray*}
\mathcal{L}_{\rm CC} &=& \frac{g}{\sqrt{2}}\, \sum_{\alpha=e, \mu, \tau}
\bigg[ \overline{\ell}_{\alpha \,L}\, \gamma_\mu {\nu}_{\alpha \,L}\, W^{\mu}_L 
      + \overline{\ell}_{\alpha \,R}\, \gamma_\mu {N}_{\alpha \,R}\, W^{\mu}_R \bigg] 
      + \text{h.c.} 
\end{eqnarray*}
}
The flavor eigenstates can be expressed in terms of mass eigenstates ($\nu_{i}$, $S_{j}$, $N_{k}$) 
as
\begin{eqnarray}
& &\nu_{\alpha\,L} \sim \mathcal{V}^{\nu \nu}_{\alpha\, i}\, \nu_{i} + \mathcal{V}^{\nu\, S}_{\alpha\, j}\, S_{j} +
                      \mathcal{V}^{\nu\, N}_{\alpha\, k}\, N_{k}\, , \nonumber \\
& &N_{\alpha\,R} \sim \mathcal{V}^{N\, \nu}_{\alpha\, i}\, \nu_{i} + \mathcal{V}^{N\, S}_{\alpha\, j}\, S_{j} +
                      \mathcal{V}^{NN}_{\alpha\, k}\, N_{k}\, ,
\end{eqnarray}
where $i,j,k=1,2,3$.
\section{Numerical estimation of $0\nu\beta\beta$ contributions}
\label{sec:numerical_estimation}
We omit a detailed discussion on fermion mass fitting at GUT scale and derivation of $M_D$, $M_N$ 
at TeV scale since this has been already done in ref.\cite{Awasthi:2013ff}. We simply use the derived 
model parameters of ref.\cite{Awasthi:2013ff} and extend the numerical estimation 
of non-standard contributions to $0\nu \beta\beta$ decay. Our major aim is to elucidate how unequal couplings 
i.e $g_{L} \neq g_{R}$ enhance the rate of $0\nu\beta\beta$ transition in $W_R-W_R$ and $W_L-W_R$ channels.  

Here we present the numerical values of all the model parameters. The Dirac neutrino mass matrix $M_D$ derived at TeV scale 
(including RG corrections) has the form \cite{Awasthi:2013ff},
\begin{eqnarray}
\label{eq:md_with_rge}
M_D =  \left( \begin{array}{ccc} 
 0.02274          & 0.09891-0.01603i    & 0.1462-0.3859i\\
 0.09891+0.01603i & 0.6319              & 4.884+0.0003034i\\
 0.1462+0.3859i   & 4.884-0.0003034i    & 117.8
\end{array} \right) \text{GeV}\, .
\end{eqnarray}
\begin{table}[h!]
\centering
\begin{tabular}{|c|c|c|c|}
\hline \hline
 EISS Scheme              & A             & B       & C \\ [2mm]
 \hline \hline
$(M_1, M_2, M_3)$        & $(30, 150, 2989)\,\mbox{GeV}$      & $(100, 100, 4877)\,\mbox{GeV}$    & $(50, 200, 1711)\,\mbox{GeV}$ \\[2mm]
\hline
$(M_{N1}, M_{N2}, M_{N3})$  & $(500, 1000, 10000)\,\mbox{GeV}$  & $(500, 5000, 10000)\,\mbox{GeV}$  & $(1250, 3000, 5000)\,\mbox{GeV}$ \\[2mm]
\hline
$m_{\nu1}$               & $0.001268\,\mbox{eV}$            & $0.001268\,\mbox{eV}$               & $0.001269\,\mbox{eV}$ \\[2mm]
\hline
$m_{\nu2}$               & $0.00879\,\mbox{eV}$            & $0.00879\,\mbox{eV}$               & $0.0088\,\mbox{eV}$ \\[2mm]
\hline
$m_{\nu3}$               & $0.049\,\mbox{eV}$            & $0.049\,\mbox{eV}$               & $0.0492\,\mbox{eV}$ \\[2mm]
\hline
$M_{S1}$               & $1.8\,\mbox{GeV}$            & $1.99\,\mbox{GeV}$               & $1.9\,\mbox{GeV}$ \\[2mm]
\hline
$M_{S2}$               & $22.05\,\mbox{GeV}$            & $2.0\,\mbox{GeV}$               & $13.27\,\mbox{GeV}$ \\[2mm]
\hline
$M_{S3}$               & $827.87\,\mbox{GeV}$            & $2000\,\mbox{GeV}$               & $532\,\mbox{GeV}$ \\[2mm]
\hline
\end{tabular}
\caption{Different mass ranges of light active neutrinos $\nu_L$, heavy RH neutrinos ($N_R$) and sterile neutrinos $S_L$ for extended inverse seesaw (EISS) scheme.}
\end{table}

The $N-S$ mixing matrix $M=\text{diag}[10.5, 120, 2500]$ GeV, and heavy RH neutrino mass matrix $M_N = \text{diag}[115,1785,7500]$ GeV 
provide mass eigenvalues to heavy sterile neutrinos $M_{S}=\text{diag}[1.06, 8.6, 887.6] \mbox{GeV}$. As an immediate outcome, 
we get two mixing matrices for 
light-light neutrinos and light-sterile neutrinos as 
\begin{eqnarray}
&&\mathcal{V}^{\nu \nu} =
\left(
\begin{array}{ccc}
 0.81494+0.00002\,i & 0.55801-0.000019\, i & -0.12659-0.091913\, i \\
-0.35953-0.049486\, i & 0.67140-0.03401\, i & 0.645155+0.00012\, i \\
 0.447078-0.057247\, i & -0.48392-0.03934\, i & 0.74554-0.000095\, i
\end{array}
\right)\, , 
\label{eq:Unu-nu}\\
&& \mathcal{V}^{\nu S}=\left(
\begin{array}{ccc}
 0.002165 & 0.00065-0.0001\, i & 0.00008-0.0002\, i \\
 0.0094 + 0.00152\, i & 0.0052  & 0.0019 + 1.16\times 10^{-7}\, i \\
 0.0139 + 0.0367\, i & 0.0406-2.49\times10^{-6}\, i & 0.0457 
\end{array}
\right)\, .
\label{eq:Unu-S}
\end{eqnarray} repectively.

Another key parameter is the mixing matrix for light and heavy RH Majorana neutrinos which is estimated to be
\begin{eqnarray}
\mathcal{V}^{\nu N}=\left(
\begin{array}{ccc}
 0.000198 & 0.000055 - 8.980\times10^{-6}\, i & 0.000019 - 0.0000514\, i \\
 0.00086 + 0.00014\, i & 0.000354 & 0.00065 + 4.\times10^{-8}\, i \\
 0.0012768 + 0.00337\, i & 0.002736 - 1.68\times10^{-7}\, i & 0.0157
\end{array}
\right)\, .
\label{eq:Unu-N}
\end{eqnarray}
\begin{eqnarray}
\mathcal{V}^{N \nu}=\left(
\begin{array}{ccc}
 0.00113266 - 0.0017284\, i & 0.00145509 - 0.00195787\, i & 0.00171826 - 0.00288543\, i \\
 0.00149292 + 7.97906\times 10^{-6}\, i & 0.00174197 + 5.26889 \times10^{-6}\, i & 0.00244734 + 1.45034\times 10^{-6}\, i \\
0.00781003 + 0.0000451934\, i & 0.0088384 + 0.0000296147\, i & 0.0130124 + 7.49586\times10^{-6}\, i
\end{array}
\right)\, .
\label{eq:UN-nu}
\end{eqnarray}
Also the mixing matrix between heavy RH Majorana neutrinos and light-sterile neutrinos can be esmitated as,
\begin{equation}
\mathcal{V}^{N N}= \text{diag}[0.995832, 0.99774, 0.94444]
\end{equation}
\begin{equation}
\mathcal{V}^{N S}= \text{diag}[-0.0913043, -0.0672269, -0.33333]
\end{equation}
Here we focus only on those scenarios where mass of neutrinos is either small or greater than the typical momentum exchange scale 
in $0\nu\beta\beta$ transition i.e $\langle p^2 \rangle \simeq \mbox{190\,MeV}^2$. The estimated effective mass parameter for 
standard contribution is found to be $0.0044$ eV for NH (Normal Hierarchy), $0.048$ eV for IH (Inverted Hierarchy) and 
$0.35$ eV for QD (Quasi Degenerate) nature of light neutrinos while the bound provided by KamLAND-Zen experiment is $0.23$ eV. A stringent limit on 
sum of the light neutrino masses by Planck collaboration i.e $\sum_{i=1}^{3} m_{\nu_i} \lesssim 0.23$ constraints 
the QD nature of light neutrinos while other hierarchical nature of light neutrinos can not be probed at present. Hence, one 
should explore possible new physics contributions in order to saturate the recent $0\nu\beta\beta$ experimental bound. We numerically 
estimate how the half-lives of the isotopes is enhanced due to different new contributions arising in the model and present a comparative study 
of the same in symmetric and asymmetric case.

Using the values of $M_D$, $M$, $M_N$ and the corresponding mixing between light active 
neutrinos $\nu_L$, heavy RH neutrinos ($N_R$) and sterile neutrinos $S$ along with other 
known parameters, the model predictions for different LNV dimensionless parameters and 
their experimental limits are presented in table \ref{tab:LNV-predic-limit}. In order to estimate these LNV parameters, we have fixed $M_{W_L}=83.187$ GeV, $M_{W_R} \gtrsim 3.1$ TeV~\cite{Aad:2015xaa,Aaboud:2018spl,Aaboud:2019wfg,Sirunyan:2018pom}, $g_{R}\simeq 0.39 - 0.632$, $g_{L}\simeq 0.632$, $m_e=0.51$ MeV and $m_p=935$ MeV. The limits on these LNV parameters have 
been derived from the recent KamLAND-ZEN experiment \cite{Ozaki:2019uyd}.

For the four diferent cases (as discussed in Appendix \ref{sec:unification}) the range for $g_R$ is tabulated in table \ref{tab:range}. However, for the calculations in the rest of the paper 
we consider three cases I, II and III since for case IIIA and IIIB the values of $ \delta = \frac{g_R}{g_L} $ are nearly equal.
\begin{table}[h!]
\centering
\begin{tabular}{|c|c|c|c|}
\hline
 Breaking Chain & $g_R$ & $g_L$ &  $ \delta = \frac{g_R}{g_L} $
           \\[2mm]
\hline \hline 
Case I & 0.632 & 0.632 & 1
       \\[2mm]
Case II & 0.589 & 0.632 & 0.93
       \\[2mm]
Case IIIA & 0.39 & 0.632 & 0.62
       \\[2mm] 
Case IIIB & 0.414 & 0.632 & 0.65
       \\[2mm] 
\hline     
\end{tabular}
\caption{Range of $g_R$ for different breaking chains}
\label{tab:range}
\end{table}

\begin{table}[h!]
\centering
\begin{tabular}{|c|c||c|}
\hline
  Effective mass parameter & Effective mass parameter &  Suppression Factor 
           \\[2mm]
  ({\bf symmetric LR model}) & ({\bf asymmetric LR model (Case II)}) & ${\large \bf m}^D_{\rm ee} {\bf \Large /} {\large \bf  m}_{\rm ee}$ 
           \\[2mm]
\hline \hline 
%           \hline
${\large \bf  m}_{\rm ee}^{\nu}=0.0044\, \mbox{eV}$   & ${\large \bf  m}_{\rm ee}^{\nu,D}=0.0044 \, \mbox{eV}$   & 1
          \\[2mm]
%           \hline
${\large \bf  m}_{\rm ee,L}^{S}=104.93\, \mbox{eV}$   & ${\large \bf  m}_{\rm ee,L}^{S,D}=104.93\, \mbox{eV}$   & 1
          \\[2mm]
%           \hline
${\large \bf  m}_{\rm ee,L}^{N}=0.0033\, \mbox{eV}$   & ${\large \bf  m}_{\rm ee,L}^{N,D}=0.0033 \, \mbox{eV}$   & 1
          \\[2mm]
%           \hline
${\large \bf  m}_{\rm ee,R}^{N}=0.040\, \mbox{eV}$   & ${\large \bf  m}_{\rm ee,R}^{N,D}=0.030 \, \mbox{eV}$   & $\left(\frac{g_R}{g_L}\right)^4 \simeq 0.75$
          \\[2mm]
%           \hline
${\large \bf  m}_{\rm ee}^{\Delta_R}= 1.74 \times 10^{-20}\, \mbox{eV}$    & ${\large \bf  m}_{\rm ee}^{\Delta_R,D}= 1.30 \times 10^{-20}\, \mbox{eV}$  & $\left(\frac{g_R}{g_L}\right)^4 \simeq 0.75$
          \\[2mm]
${\large \bf  m}_{\rm ee}^{\lambda , \nu}=1.142\, \mbox{eV}$   & ${\large \bf  m}_{\rm ee}^{\lambda, \nu,D}=0.988 \, \mbox{eV}$   & $\left(\frac{g_R}{g_L}\right)^2 \simeq 0.86$
          \\[2mm]
%           \hline
${\large \bf  m}_{\rm ee}^{\lambda , S}=0.0035\, \mbox{eV}$   & ${\large \bf  m}_{\rm ee}^{\lambda, S,D}=0.0030 \, \mbox{eV}$   & $\left(\frac{g_R}{g_L}\right)^2 \simeq 0.86$
          \\[2mm]
%           \hline
${\large \bf  m}_{\rm ee}^{\lambda , N}=4.486 \times 10^{-8}\, \mbox{eV}$   & ${\large \bf  m}_{\rm ee}^{\lambda, S,D}=3.858 \times 10^{-8}\, \mbox{eV}$   & $\left(\frac{g_R}{g_L}\right)^2 \simeq 0.86$
          \\[2mm]
%           \hline
${\large \bf  m}_{\rm ee}^{\eta}=2.16\, \mbox{eV}$   & ${\large \bf  m}_{\rm ee}^{\eta, D}=2.16 \, \mbox{eV}$   & 1
          \\[2mm]
\hline
\end{tabular}
\caption{Effective mass parameters for the present asymmetric TeV scale LR model with spontaneous D-parity breaking 
         where $SU(2)_R$ and discrete parity breaks at different scale and its comparision with those quantities in 
         symmetric LR model}
\label{tab:compare-effmassI}
\end{table}

\begin{table}[h!]
\centering
\begin{tabular}{|c|c||c|}
\hline
  Effective mass parameter & Effective mass parameter &  Suppression Factor 
           \\[2mm]
  ({\bf symmetric LR model}) & ({\bf asymmetric LR model (Case III)}) & ${\large \bf m}^D_{\rm ee} {\bf \Large /} {\large \bf  m}_{\rm ee}$ 
           \\[2mm]
\hline \hline 
%           \hline
${\large \bf  m}_{\rm ee}^{\nu}=0.0044\, \mbox{eV}$   & ${\large \bf  m}_{\rm ee}^{\nu,D}=0.0044 \, \mbox{eV}$   & 1
          \\[2mm]
%           \hline
${\large \bf  m}_{\rm ee,L}^{S}=104.93\, \mbox{eV}$   & ${\large \bf  m}_{\rm ee,L}^{S,D}=104.93\, \mbox{eV}$   & 1
          \\[2mm]
%           \hline
${\large \bf  m}_{\rm ee,L}^{N}=0.0033\, \mbox{eV}$   & ${\large \bf  m}_{\rm ee,L}^{N,D}=0.0033 \, \mbox{eV}$   & 1
          \\[2mm]
%           \hline
${\large \bf  m}_{\rm ee,R}^{N}=0.040\, \mbox{eV}$   & ${\large \bf  m}_{\rm ee,R}^{N,D}=0.0052 \, \mbox{eV}$   & $\left(\frac{g_R}{g_L}\right)^4 \simeq 0.13$
          \\[2mm]
%           \hline
${\large \bf  m}_{\rm ee}^{\Delta_R}= 1.74 \times 10^{-20}\, \mbox{eV}$    & ${\large \bf  m}_{\rm ee}^{\Delta_R,D}= 2.58 \times 10^{-21}\, \mbox{eV}$  & $\left(\frac{g_R}{g_L}\right)^4 \simeq 0.13$
          \\[2mm]
${\large \bf  m}_{\rm ee}^{\lambda , \nu}=1.142\, \mbox{eV}$   & ${\large \bf  m}_{\rm ee}^{\lambda, \nu,D}=0.411 \, \mbox{eV}$   & $\left(\frac{g_R}{g_L}\right)^2 \simeq 0.36$
          \\[2mm]
%           \hline
${\large \bf  m}_{\rm ee}^{\lambda , S}=0.0035\, \mbox{eV}$   & ${\large \bf  m}_{\rm ee}^{\lambda, S,D}=0.0013 \, \mbox{eV}$   & $\left(\frac{g_R}{g_L}\right)^2 \simeq 0.37$
          \\[2mm]
%           \hline
${\large \bf  m}_{\rm ee}^{\lambda , N}=4.486 \times 10^{-8}\, \mbox{eV}$   & ${\large \bf  m}_{\rm ee}^{\lambda, S,D}=1.615 \times 10^{-8}\, \mbox{eV}$   & $\left(\frac{g_R}{g_L}\right)^2 \simeq 0.36$
          \\[2mm]
%           \hline
${\large \bf  m}_{\rm ee}^{\eta}=2.16\, \mbox{eV}$   & ${\large \bf  m}_{\rm ee}^{\eta, D}=2.16 \, \mbox{eV}$   & 1
          \\[2mm]
\hline
\end{tabular}
\caption{Effective mass parameters for the present asymmetric TeV scale LR model with spontaneous D-parity breaking (after introducing Pati-Salam symmetry) 
         where $SU(2)_R$ and discrete parity breaks at different scale and its comparision with those quantities in 
         symmetric LR model}
\label{tab:compare-effmassII}
\end{table}

\begin{table}[h!]
\centering
\begin{tabular}{|c|c|c||c|}
\hline \hline
Nucleus 
                  & Model prediction for ${\large \bf T}_{1/2}^{0\nu}$ (yrs) 
                                     &  Current Limits & Future Limits  \\[2mm]
\hline \hline 
$^{76}\mbox{Ge}$  & $1.3\times 10^{25}$-$4.13\times 10^{27}$
                                     & $\gtrsim 8.0 \times 10^{25}$ yrs 
                                     & $\gtrsim 6.0 \times 10^{27}$ yrs  \\[2mm]
\hline
$^{136}\mbox{Xe}$  & $1.03\times 10^{25}$-$6.13\times 10^{27}$
                                     & $\gtrsim 1.6 \times 10^{26}$ yrs
                                     & $\gtrsim 8.0 \times 10^{27}$ yrs \\[2mm]
\hline
\end{tabular}
\caption{Estimated value of half-life for $0\nu\beta\beta$ transition due to different nuclei and 
         corresponding experimental limits.}
\label{tab:esthlife-limit}
\end{table}

%****************************************************************************
\begin{table}[h!]
\centering
\begin{tabular}{|c|c|c|c|c|}
\hline \hline
 LNV    & Estimated value  & Estimated value   & Estimated value  &   Experimental 
 \\[2mm]
  parameters   &  (\bf Case I)    &  (\bf Case II)   &  (\bf Case III) &      Limit   \\[2mm]
\hline \hline
$\eta_{\nu}$    & $8.73 \times 10^{-9}$ & $8.73 \times 10^{-9}$  & $8.73 \times 10^{-9}$  & $\lesssim 2.66 \times 10^{-7}$ \\[2mm]
\hline
$\eta^{S,N}_{LL}$   & $2.45 \times 10^{-10}$ & $2.45 \times 10^{-10}$  & $2.45 \times 10^{-10}$        & $\lesssim 2.55 \times 10^{-9}$  \\[2mm]
\hline
 $\eta^{\nu}_{RR}$  & $1.932 \times 10^{-19}$ & $1.445 \times 10^{-19}$ & $2.50 \times 10^{-20}$       & $\lesssim 2.66 \times 10^{-7}$  \\[2mm]
\hline
$\eta^{S,N}_{RR}$   & $1.09 \times 10^{-9}$ & $ 8.154 \times 10^{-10}$  & $1.41 \times 10^{-10}$        & $\lesssim 2.55 \times 10^{-9}$  \\[2mm]
\hline
$\eta^{\Delta_R}_{RR}$   & $ 1.63 \times 10^{-11}$ & $ 1.22 \times 10^{-11}$  & $ 2.41 \times 10^{-12}$        & $\lesssim 2.55 \times 10^{-9}$  \\[2mm]
\hline
$\eta^{}_{\eta}$         & $2.65 \times 10^{-9}$  & $2.65 \times 10^{-9}$  & $2.65 \times 10^{-9}$        & $\lesssim 1.13 \times 10^{-9}$  \\[2mm]
\hline
$\eta^{}_{\lambda}$      & $1.51 \times 10^{-7}$ & $1.31 \times 10^{-7}$  & $5.43 \times 10^{-8}$        & $\lesssim 2.18 \times 10^{-7}$ \\[2mm]
\hline
\end{tabular}
\caption{Estimated values of lepton number violating (LNV) parameters within the framework of 
TeV scale asymmetric left-right model with $g_{L}\neq g_{R}$.}
\label{tab:LNV-predic-limit}
\end{table}

\section{Comparative study of $0\nu\beta\beta$ contributions with D-parity breaking mechanism}
\label{sec:comparative_study}
Predictions on neutrinoless double beta decay in LR model with Spontaneous D-parity breaking and half-life with proper nuclear matrix element and normalized lepton number violating effective mass
parameters have been discussed in the appendix section. In this section we present a comparative study of different contributions to neutrinoless double beta decay process 
arising due to mediation of either one $W^-_R$ or two $W^-_R$ gauge bosons in terms of half-life and effective mass parameters 
within the frameworks of symmetric and asymmetric left-right model. In an asymmetric left-right model, 
when we express non-standard contributions for $0\nu\beta\beta$ via $W_R-W_R$ and $W_L-W_R$ chanels which involves $SU(2)_R$ gauge coupling 
transitions in terms of known parameters like Fermi coupling constant $G^2_F$ then these expressions carry a overall factor 
like $\frac{g_{R}}{g_{L}}$ in Feynman amplitudes and $\frac{g_{L}}{g_{R}}$ in half-life estimation. Depending upon the value of $g_{L}$, $g_{R}$ and 
the ratio $\frac{g_{R}}{g_{L}}$, there are different contributions which are either suppressed or enhanced 
as compared to the $0\nu\beta\beta$ transition derived in a symmetric left-right theory. 

We have presented the numerical estimation of effective mass parameters compairing the symmetric (case-I) as well as asymmetric (case-II) LR model (without introducing Pati Salam symmetry) in table \ref{tab:compare-effmassI}. For a non-SUSY $SO(10)$ GUT where Pati-Salam symmetry with D-parity occurs as the highest intermediate symmetry breaking step,
the RG evolution of gauge couplings predicts $g_{L}=0.65$ and $g_{R}=0.39$ at TeV scale. This particular set 
up gives the ratio $\left(\frac{g_{R}}{g_{L}}\right)=0.6$. The numerical values of effective mass parameter in symmetric and asymmetric 
cases are compared in table \ref{tab:compare-effmassII}. The suppresion factor in effective mass parameters is found to be 
$\left( \frac{g_{R}}{g_{L}} \right)^4\simeq 0.13$ in the $W_R-W_R$ chanel via exchange of heavy RH neutrinos 
and heavy RH Higgs triplets while in the $W_L-W_R$ channel it is found to be $\left( \frac{g_{R}}{g_{L}} \right)^2\simeq 0.36$. 
Similarly, new contributions involving right-handed charged current interaction are enhanced and the enhancement factor is found to be 
$\left( \frac{g_{L}}{g_{R}} \right)^8\simeq 59.29$ in the $W_R-W_R$ chanel via exchange of heavy RH neutrinos and RH Higgs triplets while it is 
$\left( \frac{g_{L}}{g_{R}} \right)^2\simeq 7.7$ in the $W_L-W_R$ channel. The same is shown in table \ref{tab:compare-half-lifeII}. If we do not introduce the Pati Salam symmetry in the unification scenario, we will get negligible enhancement of half-life estimation for asymmetric LR model as compared to symmetric one. The result of such estimation has been tabulated in table \ref{tab:compare-half-lifeI}. 

\begin{table}[h!]
\centering
\begin{tabular}{|c|c||c|}
\hline
  Half-life & Half-life &  Enhancement Factor 
           \\[2mm]
  ({\bf Case I}) & ({\bf Case II}) & $\left[{\large \bf T}_{1/2}^{0\nu}\right]_{D} {\bf \Large /}
                                    \left[{\large \bf T}_{1/2}^{0\nu}\right]$ 
           \\[2mm]
\hline \hline 
$\left[{\large \bf T}_{1/2}^{0\nu}\right]_{N}={\bf \Large 1/}\left(\mathcal{K}_{0\nu}|{\large \bf m}_{\rm ee}^{N}|^2\right)$   
        & $\left[{\large \bf T}_{1/2}^{0\nu}\right]_{N,D}={\bf \Large 1/}\left(\mathcal{K}_{0\nu}|{\large \bf m}_{\rm ee,D}^{N}|^2\right)$   
                         & $\left(\frac{g_L}{g_R}\right)^8 \simeq 1.78$
          \\[3mm]
% \hline
$\left[{\large \bf T}_{1/2}^{0\nu}\right]_{\Delta_R}={\bf \Large 1/}\left(\mathcal{K}_{0\nu}|{\large \bf m}_{\rm ee}^{\Delta_R}|^2\right)$   
       & $\left[{\large \bf T}_{1/2}^{0\nu}\right]_{\Delta_R,D}={\bf \Large 1/}\left(\mathcal{K}_{0\nu}|{\large \bf m}_{\rm ee,D}^{\Delta_R}|^2\right)$  
                         & $\left(\frac{g_L}{g_R}\right)^8 \simeq 1.78$
          \\[3mm]
$\left[{\large \bf T}_{1/2}^{0\nu}\right]_{\lambda}={\bf \Large 1/}\left(\mathcal{K}_{0\nu}|{\large \bf m}_{\rm ee}^{\lambda}|^2\right)$   
       & $\left[{\large \bf T}_{1/2}^{0\nu}\right]_{\lambda,D}={\bf \Large 1/}\left(\mathcal{K}_{0\nu}|{\large \bf m}_{\rm ee,D}^{\lambda}|^2\right)$  
                         & $\left(\frac{g_L}{g_R}\right)^4 \simeq 1.33$
          \\[3mm]
\hline
\end{tabular}
\caption{Expression for half-lives governing $0\nu\beta\beta$ transition in TeV scale symmetric and asymmetric left-right models. 
         $\left[{\large \bf T}_{1/2}^{0\nu}\right]_{D}$ stands for half-life expression in case of a LR model 
         with spontaneous D-parity breaking while $\left[{\large \bf T}_{1/2}^{0\nu}\right]$ stands
         for half-life expression in symmetric LR model.}
\label{tab:compare-half-lifeI}
\end{table}

\begin{table}[h!]
\centering
\begin{tabular}{|c|c||c|}
\hline
  Half-life & Half-life &  Enhancement Factor 
           \\[2mm]
  ({\bf Case I}) & ({\bf Case III}) & $\left[{\large \bf T}_{1/2}^{0\nu}\right]_{D} {\bf \Large /}
                                    \left[{\large \bf T}_{1/2}^{0\nu}\right]$ 
           \\[2mm]
\hline \hline 
$\left[{\large \bf T}_{1/2}^{0\nu}\right]_{N}={\bf \Large 1/}\left(\mathcal{K}_{0\nu}|{\large \bf m}_{\rm ee}^{N}|^2\right)$   
        & $\left[{\large \bf T}_{1/2}^{0\nu}\right]_{N,D}={\bf \Large 1/}\left(\mathcal{K}_{0\nu}|{\large \bf m}_{\rm ee,D}^{N}|^2\right)$   
                         & $\left(\frac{g_L}{g_R}\right)^8 \simeq 59.29$
          \\[3mm]
% \hline
$\left[{\large \bf T}_{1/2}^{0\nu}\right]_{\Delta_R}={\bf \Large 1/}\left(\mathcal{K}_{0\nu}|{\large \bf m}_{\rm ee}^{\Delta_R}|^2\right)$   
       & $\left[{\large \bf T}_{1/2}^{0\nu}\right]_{\Delta_R,D}={\bf \Large 1/}\left(\mathcal{K}_{0\nu}|{\large \bf m}_{\rm ee,D}^{\Delta_R}|^2\right)$  
                         & $\left(\frac{g_L}{g_R}\right)^8 \simeq 59.29$
          \\[3mm]
$\left[{\large \bf T}_{1/2}^{0\nu}\right]_{\lambda}={\bf \Large 1/}\left(\mathcal{K}_{0\nu}|{\large \bf m}_{\rm ee}^{\lambda}|^2\right)$   
       & $\left[{\large \bf T}_{1/2}^{0\nu}\right]_{\lambda,D}={\bf \Large 1/}\left(\mathcal{K}_{0\nu}|{\large \bf m}_{\rm ee,D}^{\lambda}|^2\right)$  
                         & $\left(\frac{g_L}{g_R}\right)^4 \simeq 7.7$
          \\[3mm]
\hline
\end{tabular}
\caption{Expression for half-lives governing $0\nu\beta\beta$ transition in symmetric and asymmetric left-right models. Case-I is for 
symmetric left-right model ($g_L=g_R$), Case-III is where D-parity breaking in Pati-Salam symmetry leads to $g_L\neq g_R$.}
\label{tab:compare-half-lifeII}
\end{table}

\section{Results and Conclusion}
\label{sec:conclusion}
\begin{figure}[h]
\centering
\includegraphics[scale=0.5]{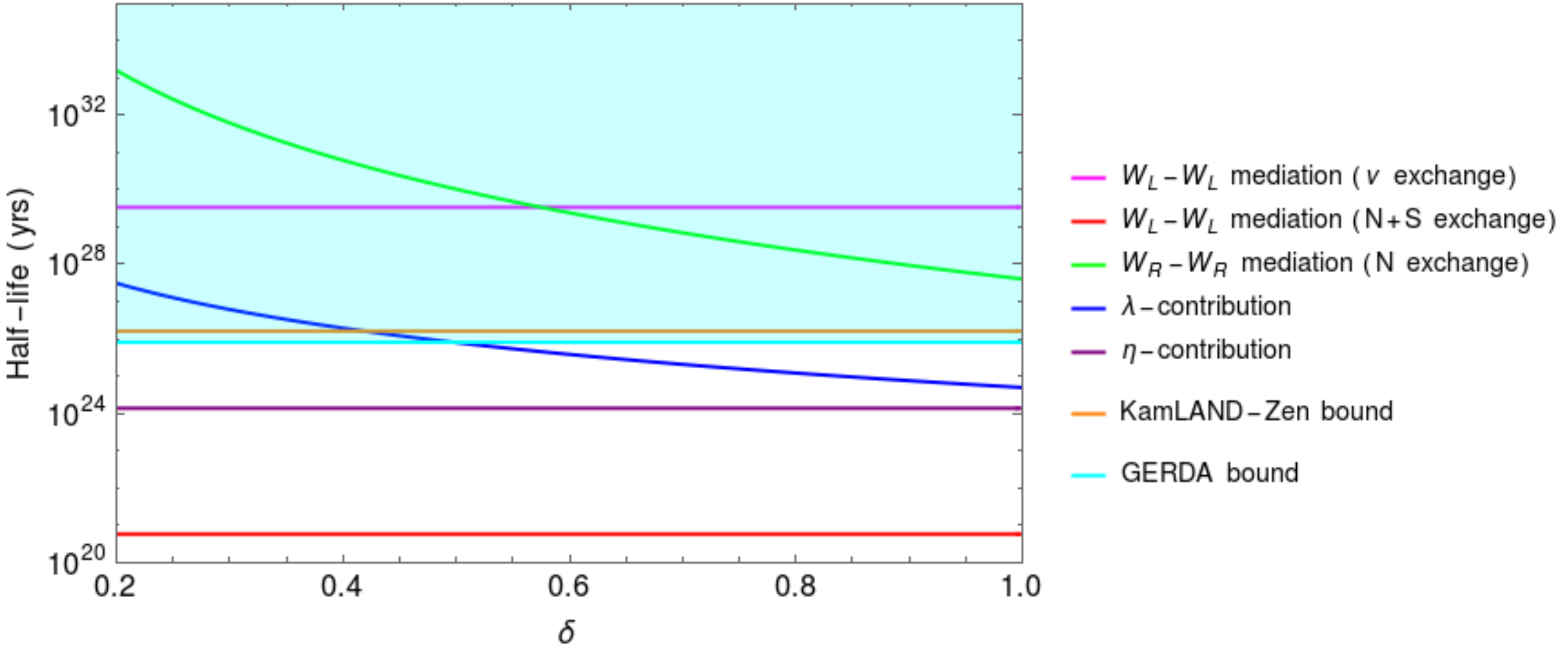}
\caption{Half life of $0\nu\beta\beta$ process due to all possible channels in the model vs $\delta$ ($=\frac{g_R}{g_L}$). The orange horizontal line 
represents recent KamLAND-Zen bound while the blue line represents GERDA bound on half life of the process. The cyan shaded region shows the allowed range for 
half life.}
\label{hlife_delta}
\end{figure}
\begin{figure}[h]
\centering
\includegraphics[scale=1]{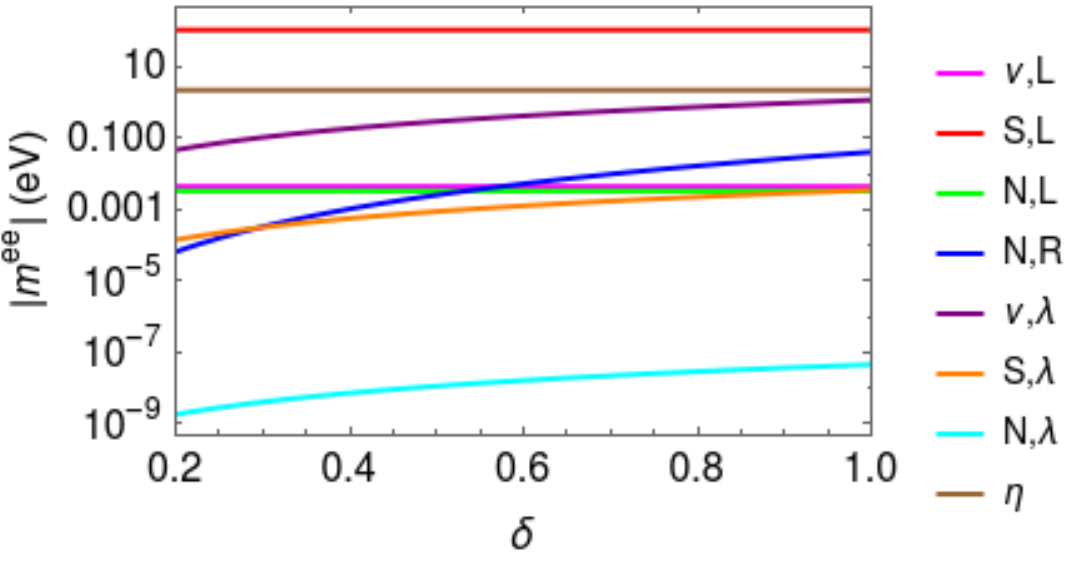}
\caption{The plot shows how effective Majorana mass parameter due to different decay channels varies with $\delta$. }
\label{mee_delta}
\end{figure}
\begin{figure}[h]
\centering
\includegraphics[scale=0.37]{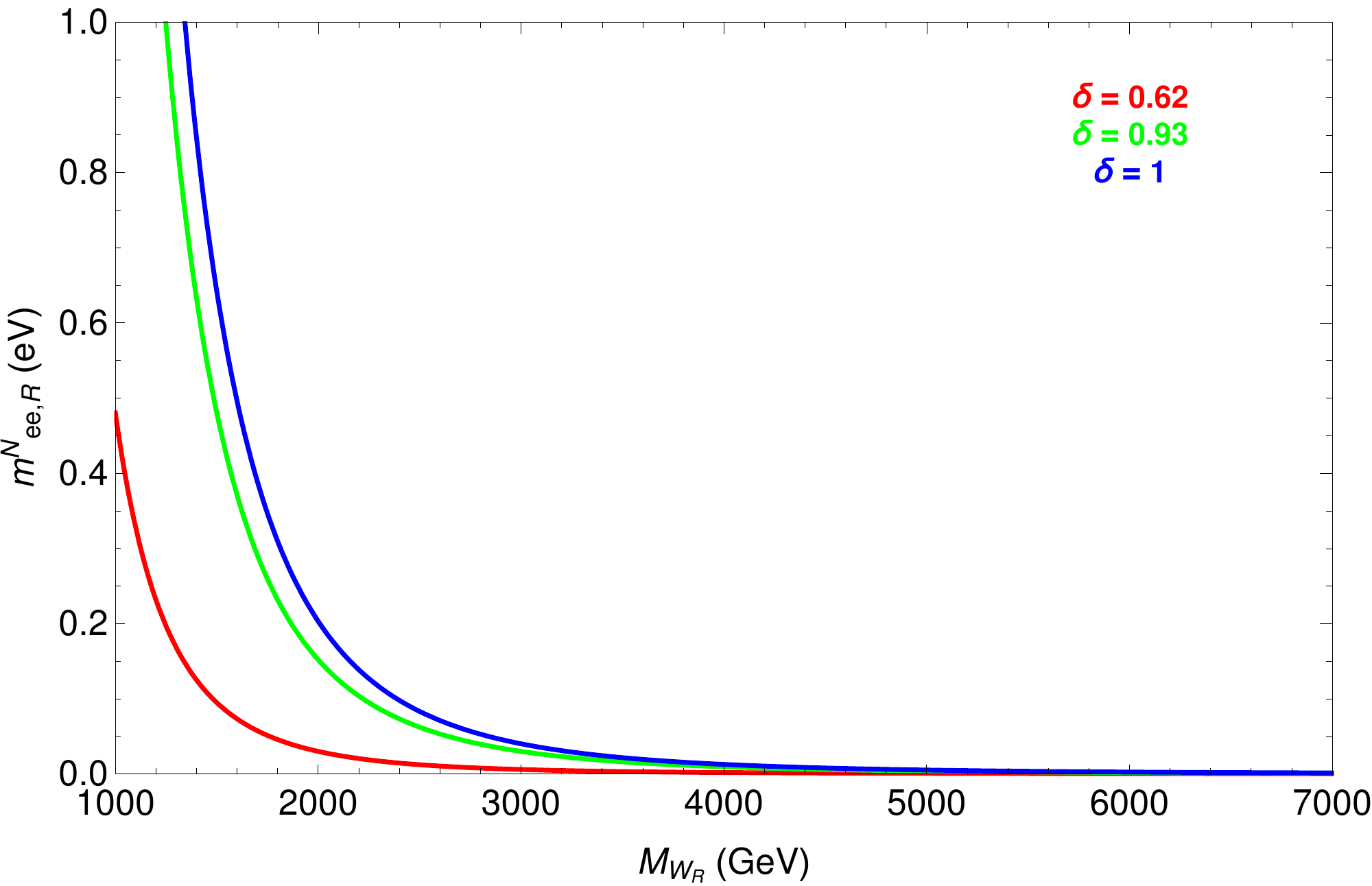}\qquad
\includegraphics[scale=0.37]{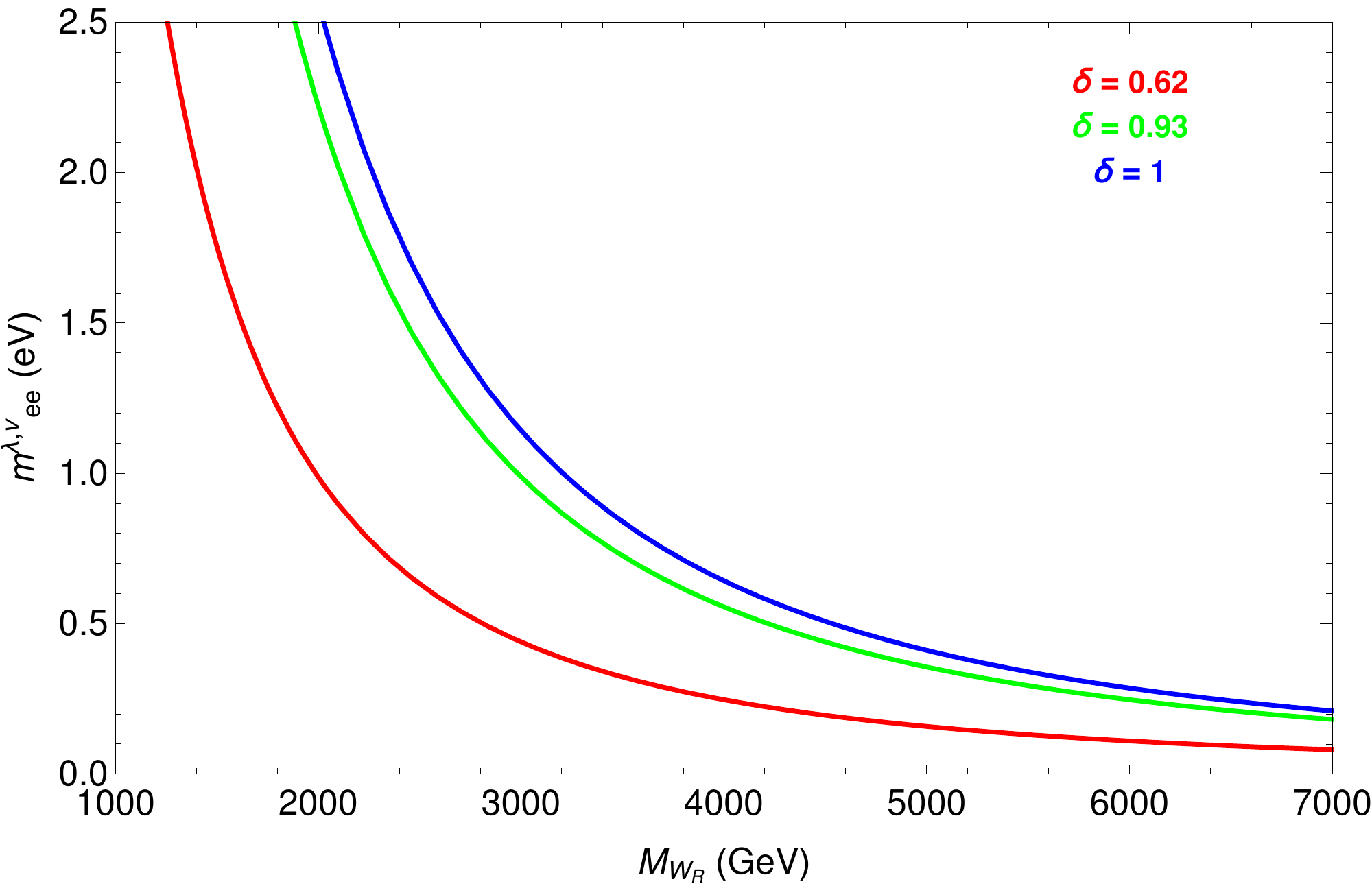}
\caption{The plot in the left panel shows effective majorana mass parameter due to heavy neutrino, $N$ exchange in purely 
right-handed currents vs $W_R$ mass, where $W_R$ mass varies from 1 to 3 TeV. 
The plot in the right panel shows effective majorana mass parameter due to $W_L-W_R$ mixing ($\lambda$ diagram) with $\nu$ exchange vs $W_R$ mass.
In both the plots three different values of $\delta$ are considered: $\delta$=0.63, 0.93, 1.}
\label{mee_WR}
\end{figure}
\begin{figure}[h]
\centering
\includegraphics[scale=0.37]{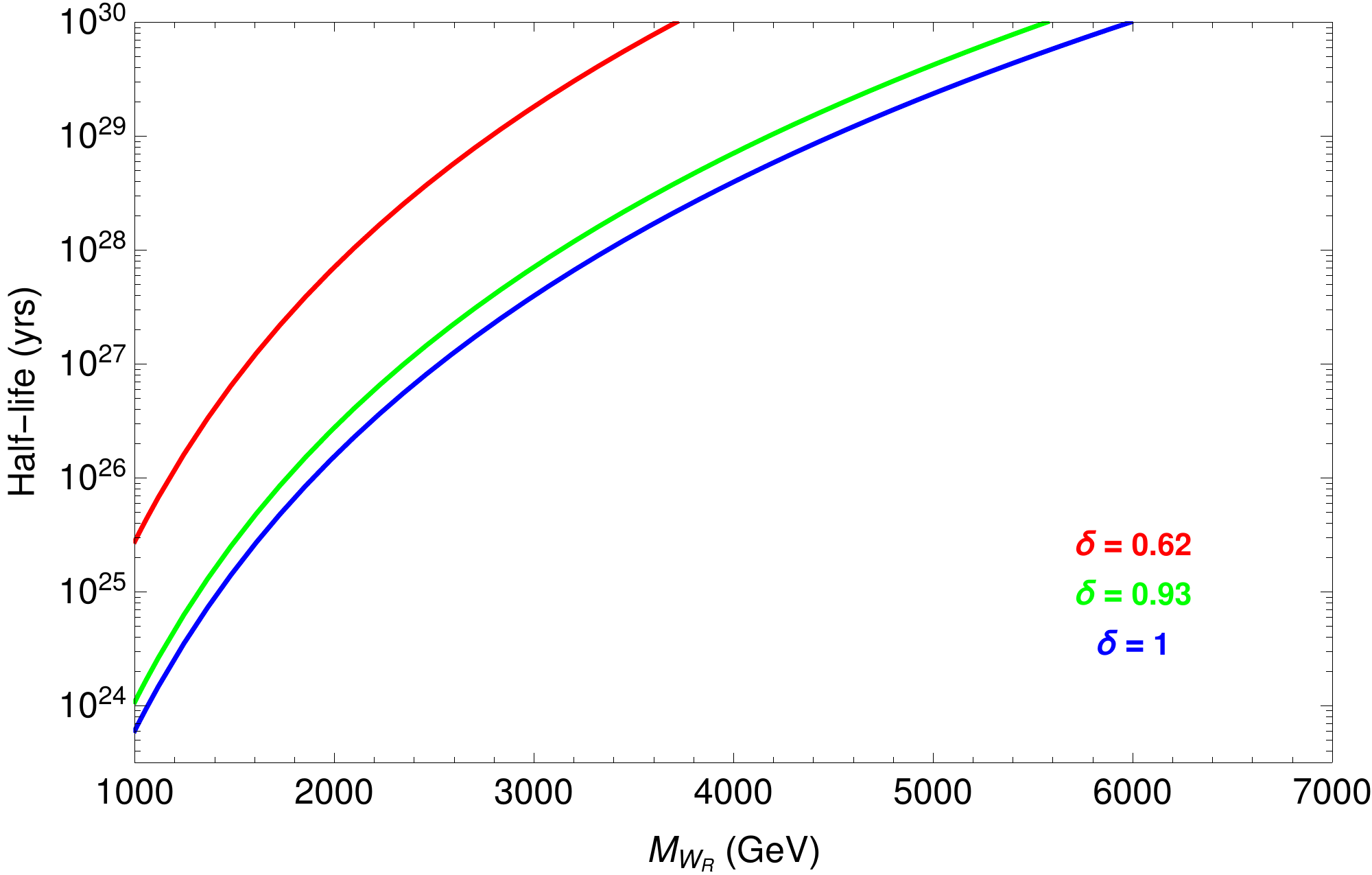}\qquad
\includegraphics[scale=0.37]{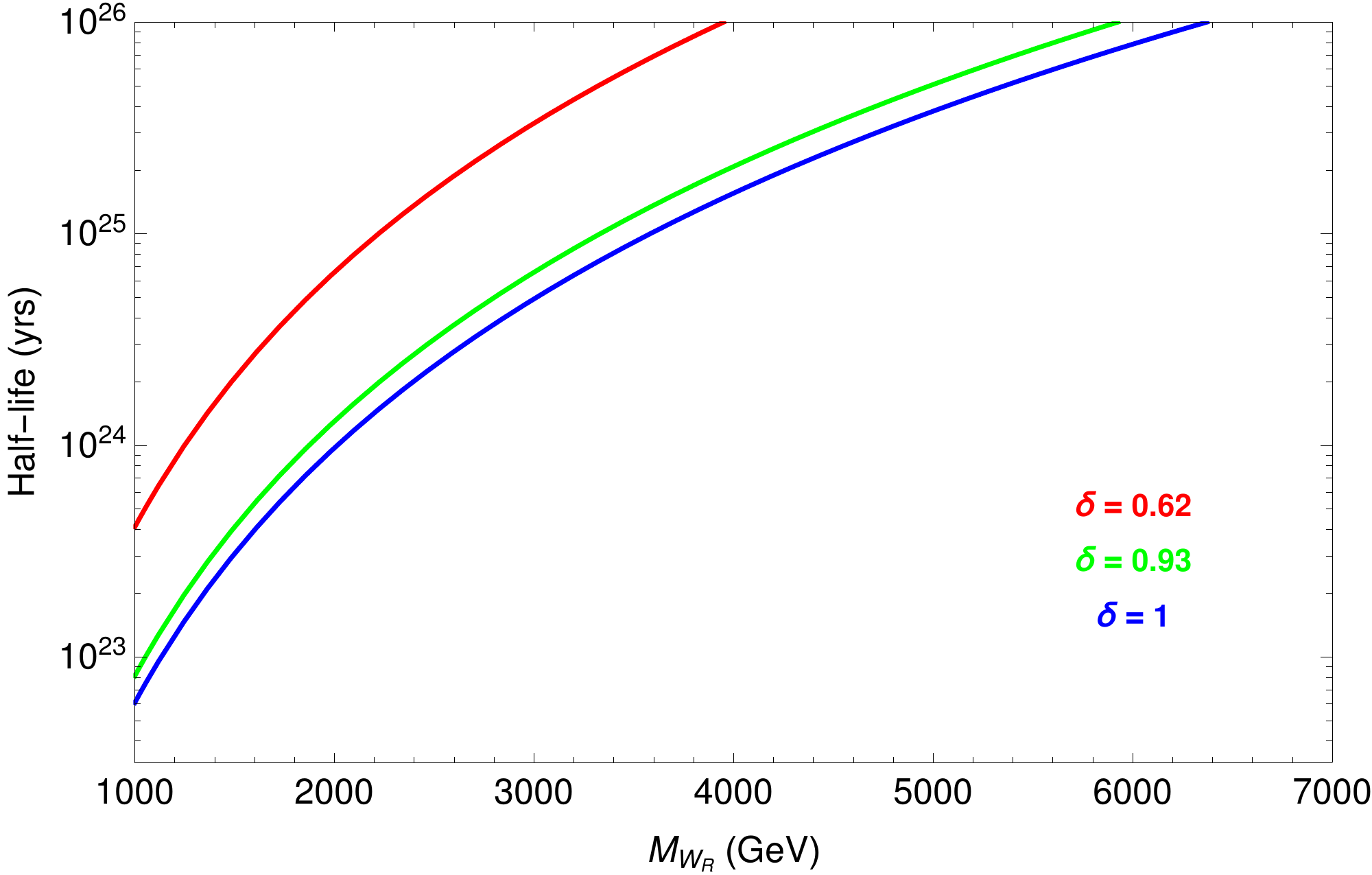}
\caption{Plot in the left panel shows half life due to $N$ exchange in $W_R-W_R$ channel vs mass of $W_R$ while plot in the right panel shows 
half life due to all $\lambda$ diagrams ($\nu,N,S$ exchange with $W_L-W_R$ mixing) vs mass of $W_R$. In both the plots three different values of $\delta$ are considered: $\delta$=0.63, 0.93, 1.}
\label{hlife_WR}
\end{figure}
One important outcome of extended inverse seesaw scheme is that type-I seesaw contribution 
is exactly canceled out thereby allowing the possibility of large left-right mixing in the 
neutrino sector. The occurance of Pati-Salam symmetry at the highest scale gives large value to Dirac neutrino 
mass matrix $M_D$ and thus the mixed helicity $\lambda$ and $\eta$ diagrams contribute dominantly to the 
$0\nu\beta\beta$ transition. At the same time, the $W_R-W_R$ mediated diagrams due to exchange of heavy RH neutrinos 
also deliver dominant contributions to the process. In addition, another important contribution 
comes from purely left-handed currents via $W_L-W_L$ mediation due to exchange of heavy sterile neutrinos. 
We have also discussed that LR model with spontaneous D-parity breaking mechanism gives 
different analytic expressions for different $0\nu\beta\beta$ contributions in the $W_R-W_R$ and $W_L-W_R$ 
mediated channels since the theory predicts unequal gauge couplings for $SU(2)_L$ and $SU(2)_R$ 
gauge groups. 
We have embedded the model in non-supersymmetric $SO(10)$ GUT and discussed different symmetry breaking chains, 
i.e. with and without Pati-Salam symmetry for showing how the enhancement factor $\left(\frac{g_{L}}{g_{R}} \right)$ 
for half-life prediction of neutrinoless double beta decay changes for different cases.  
When Pati-Salam symmetry is not included in the symmetry breaking chain, we get the enhancement factor 
$\left(\frac{g_{L}}{g_{R}} \right)^8\simeq 1.78$ for $W_R-W_R$ channel while for the $W_L-W_R$ channel 
the enhancement factor is $\left(\frac{g_{L}}{g_{R}} \right)^4\simeq 1.33$. 
However the enhancement factor increases significantly when Pati-Salam symmetry appears in the symmetry breaking chain.
In this case, the enhancement factor becomes $\left(\frac{g_{L}}{g_{R}} \right)^8\simeq 59.29$ for $W_R-W_R$ channel 
and for $W_L-W_R$ channel the enhancement factor becomes $\left(\frac{g_{L}}{g_{R}} \right)^4\simeq 7.7$.
Pati-Salam symmetry also plays an important role in predicting values of $SU(2)_{L}$ and $SU(2)_{R}$ 
gauge couplings for which we get the values; $g_{L}=0.632$ and $g_{R}\simeq 0.39$. 

We have shown various plots to infer how half-life of $0\nu\beta\beta$ decay due to different channels varies 
with the ratio $\frac{g_R}{g_L}$ i.e. $\delta$ and mass of $W_R$. From Fig.\ref{hlife_delta} we see that the 
cyan shaded region is sensitive to the current KamLAND-Zen and GERDA bounds. Since the value of the ratio $\frac{g_R}{g_L}$ 
ranges from $0.62$ to $1$ for three different cases considered in the work, the plot shows only the contributions 
from $W_L-W_L$ channel due to light neutrino exchange and from $W_R-W_R$ channel due to heavy neutrino exchange lie 
within the priviledged region. Fig.\ref{mee_delta} shows how the effective Majorana mass parameter varies with the ratio $\delta$. Only non-trivial $M_{W_R}$ dependence occurs for the contribution arise from $W_R-W_R$ mediation with RH neutrino exchange as well as from $W_L-W_R$ mediation ($\lambda$-contribution). So, 
Fig.\ref{mee_WR} shows how the effective Majorana mass parameter due to these two decay channels vary with the mass of $W_R$ 
and the variation of half life with $W_R$ mass has been presented in Fig.\ref{hlife_WR}. For Fig.~\ref{mee_WR} and \ref{hlife_WR}, the mass range for $W_R$ has been considered here as, $M_{W_R} \in [1,7]$~TeV for better transperancy.  

\section{Acknowledgements}

SS is thankful to UGC for fellowship grant to support her research work. The authors thank Prof. Urjit A. Yajnik for useful comments and discussion.\\

%\hrule
\appendix
\section{Predictions on neutrinoless double beta decay in LR model with Spontaneous D-parity breaking}
\label{sec:a1}

The importance of neutrinoless double beta decay process in particle physics is far-reaching in the sense that it is one such process which can 
confirm the Majorana nature of neutrino and also provide information about the absolute scale of light neutrino mass.
Neutrinoless Double Beta Decay can be induced by the exchange of a light Majorana neutrino, which is called the standard mechanism or 
by some other lepton number violating physics which is called the non-standard 
interpretation~\cite{Mohapatra:1986su,Babu:1995vh,Hirsch:1995vr,Hirsch:1995ek,Deppisch:2012nb,Humbert:2015yva,Pas:2000vn,Deppisch:2006hb,Pas:2015eia,Mitra:2011qr,Pritimita:2016fgr,Cirigliano:2018yza,Cirigliano:2017djv} .
In the standard mechanism the parent nucleus emits a pair of virtual $W$ bosons, and
then these exchange a Majorana neutrino to produce the outgoing electrons. At the vertex where it
is emitted, the exchanged neutrino is created, in association with an electron, as an antineutrino with
almost total positive helicity, and only its small, $O(m_\nu /E)$, negative-helicity component is absorbed
at the other vertex. In LRSM the process can be mediated by heavy right-handed neutrino and some new channels can also appear due to 
left-right mixing,i.e. $W_L-W_R$ mixing. In the considered model many diagrams are possible due to the presence of heavy neutrinos $S, N$, 
doubly charged higgs scalar $\Delta_R$ and $W_L-W_R$ mixing. We will discuss that in this section, but we start by writing the 
charged current interaction Lagrangian for the model in flavor basis. 
\begin{eqnarray}
\mathcal{L}_{\rm CC} &=& \frac{g_{L}}{\sqrt{2}}\, \sum_{\alpha=e, \mu, \tau} \overline{\ell}_{\alpha \,L}\, \gamma_\mu {\nu}_{\alpha \,L}\, W^{\mu}_L 
    + \frac{g_{R}}{\sqrt{2}}\, \sum_{\alpha=e, \mu, \tau} \overline{\ell}_{\alpha \,R}\, \gamma_\mu {N}_{\alpha \,R}\, W^{\mu}_R + \text{h.c.} 
      \nonumber \\
&=& \frac{g_{L}}{\sqrt{2}}\,
\overline{e}_{\,L}\, \gamma_\mu {\nu}_{e \,L}\, W^{\mu}_L  
   + \frac{g_{R}}{\sqrt{2}}\,\overline{e}_{\,R}\, \gamma_\mu {N}_{e \,R}\, W^{\mu}_R + \text{h.c.} \quad \mbox{only for first generation}
     \nonumber \\
&=& \frac{g_{L}}{\sqrt{2}}\,
\bigg[ \overline{e}_{\,L}\, \gamma_\mu 
        \{\mathcal{V}^{\nu\nu}_{e\, i}\, \nu_i + \mathcal{V}^{\nu\, S}_{e\, i}\, S_i +
                      \mathcal{V}^{\nu\, N}_{e\, i}\, N_i \}\, 
              W^{\mu}_L \bigg] +\mbox{h.c.} \nonumber \\
    & &  + \frac{g_{R}}{\sqrt{2}}\,
\bigg[ \overline{e}_{\,R}\, \gamma_\mu 
      \{\mathcal{V}^{N\, \nu}_{\alpha\, i}\, \nu_i + \mathcal{V}^{N\, S}_{\alpha\, i}\,S_i +
                      \mathcal{V}^{NN}_{\alpha\, i}\, N_i \}\, 
              W^{\mu}_R \bigg] + \mbox{h.c.}
\label{eqn:ccint-flavor}
\end{eqnarray}
Since we have considered that the left-handed and right-handed charged gauge bosons mix with each other the physical gauge bosons 
can be expressed as a linear combinations of $W_L$ and $W_R$ as,
\begin{equation}
\label{eqn:LRmix}
\left\{ 
\begin{array}{l} 
W_1 = \phantom{-} \cos\xi ~W_L + \sin\xi ~W_R \\ 
W_2 = - \sin\xi ~W_L +  \cos\xi ~W_R 
\end{array} \right. 
\end{equation}
with mixing angle $\xi$, we have
\begin{equation}
|\tan\, 2\xi| \sim \frac{k_1\, k_2}{v^2_R} \sim \frac{k_2}{k_1} \frac{g^2_{\rm R}}{g^2_{\rm L}} 
\left (\frac{M^2_{W_L}}{M^2_{W_R}} \right) \leq 10^{-4}.
\end{equation}
\begin{figure}[htb!]
\centering
\includegraphics[scale=0.55,angle=0]{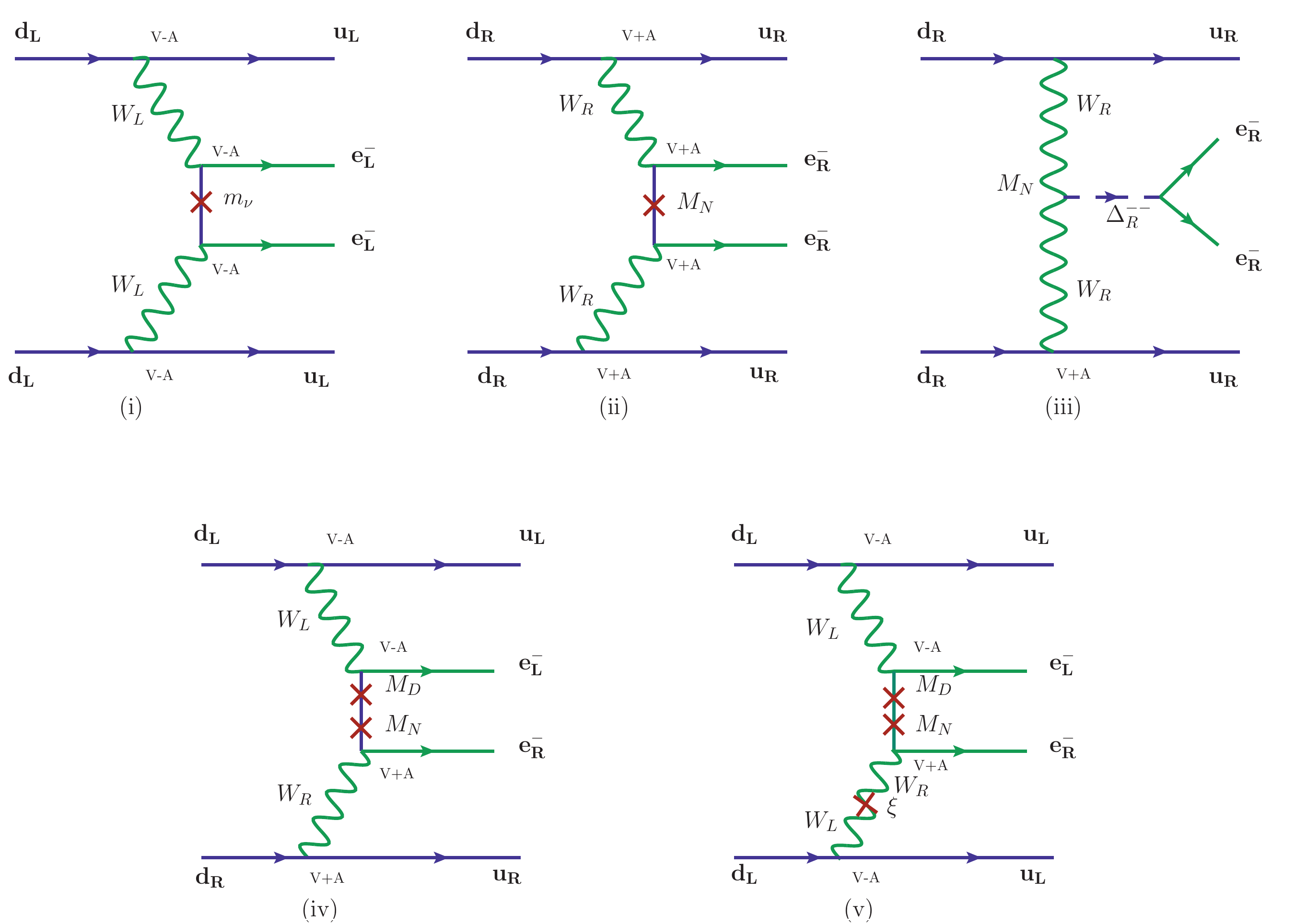}
\caption{Relevant Feynman diagrams contributing to neutrinoless double decay process 
within the framework of left-right symmetric models.}
\label{Feyn:all_LR}
\end{figure}
Different types of Feynman diagrams contributing to the $0\nu\beta\beta$ process are \cite{Awasthi:2013ff} shown in Fig \ref{Feyn:all_LR}. 
\begin{enumerate}
\item[${\bf (i)}$] Feynman diagrams in $W_L^--W_L^-$ channel (with two left-handed currents)\,
\item[${\bf (ii)}$] Feynman diagrams in $W_R^--W_R^-$ channel (with two right-handed currents)\,
\item[${\bf (iii)}$] Doubly charged Higgs scalar exchange with right-handed currents (this can also be possible with left-handed currents)\,
\item[${\bf (iv)}$] Neutrino and $W_R$ exchanges with Dirac mass helicity flip in $W_L-W_R$ channel ($\lambda$ mechanism)\,
\item[${\bf (v)}$] Neutrino and $W_L$ exchanges with Dirac mass helicity flip and $W_L-W_R$ mixing in the $W_L-W_R$ channel ($\eta$ mechanism)\,       
\end{enumerate}
\subsection{Mass-dependent mechanisms Due to $W_L^--W_L^-$ channel and $W_R^--W_R^-$ channel}
\begin{figure}[htb!]
\centering
\includegraphics[scale=0.59,angle=0]{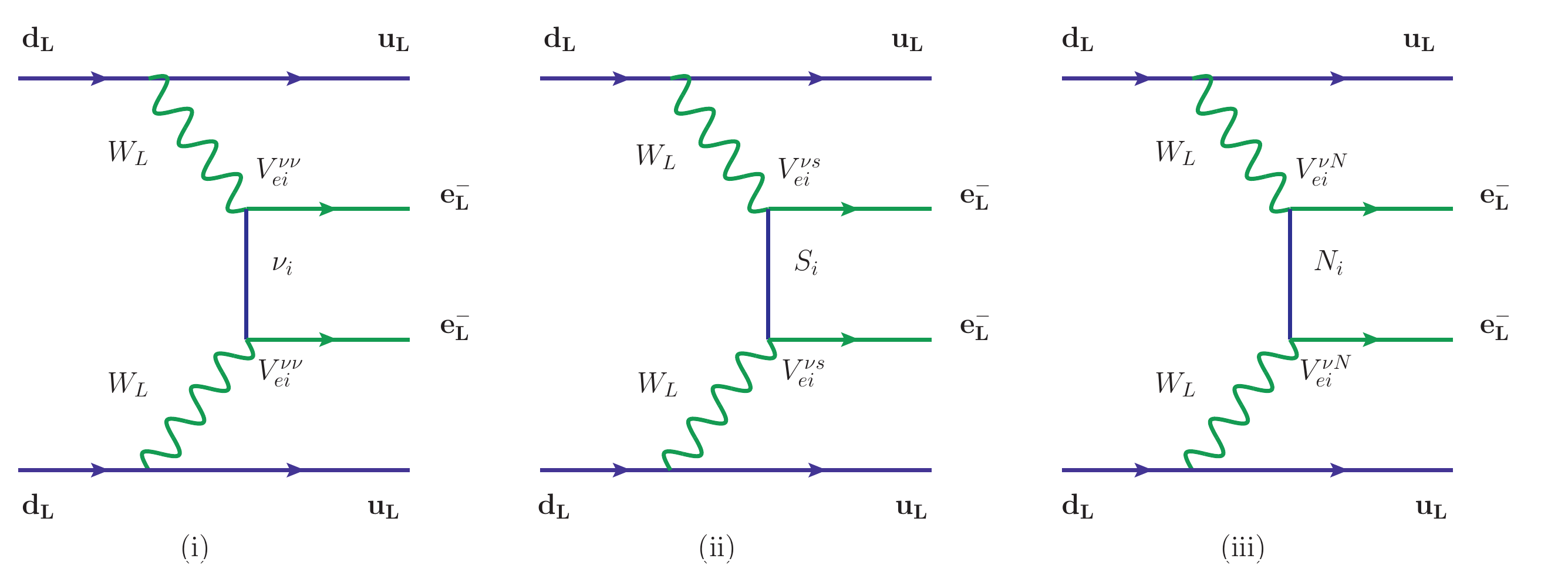}
\caption{Relevant Feyman Diagrams due to $W_L^--W_L^-$ channel}
\label{Feyn:wl_wl}
\end{figure}

Now, let's write the amplitudes for these processes and the corresponding particle physics parameter involving lepton number violation.
The Feynman amplitude for the processes having both left-handed electrons is proportional to
\begin{equation}
 {\cal A}_{LL} \simeq G_F^2\left(1+2\tan\xi+\tan^2\!\xi\right)\sum_i\left(\frac{{\mathcal{V}^{\nu\nu}_{e\,i}}^2 m_i}{p^2}
 -\frac{{\mathcal{V}^{\nu S}_{e\,i}}^2}{M_{S_i}} -\frac{{\mathcal{V}^{\nu N}_{e\,i}}^2}{M_{N_i}} \right)\, .
\label{eq:ll_amplitudes}
\end{equation}

where, $m_i$, $M_{S_i}$, and $M_{N_i}$ are the masses of light neutrino $\nu$ and heavy neutrinos S and N respectively and $\tan \xi$ represents 
left-right gauge boson mixing. The diagrams are separately shown in Fig \ref{Feyn:wl_wl}.

Similarly, the Feynman amplitudes for the processes involving $W_R^--W_R^-$ mediation via exchanges 
of either light or heavy neutrinos where both the emitted electrons are right-handed is proportional to,
\begin{eqnarray}
& &{\cal A}_{RR} \simeq G_F^2\bigg[\left(\frac{\mwl}{\mwr}\right)^4\, \left(\frac{g_{R}}{g_{L}}\right)^4 
                                 +2\left(\frac{\mwl}{\mwr}\right)^2\,\left(\frac{g_{R}}{g_{L}}\right)^2 \tan\xi
                                 +\tan^2\!\xi\bigg] \nonumber \\
&&\hspace*{4cm} \times
              \sum_i\left(\frac{{\mathcal{V}^{\nu N}_{e\,i}}^2 m_i}{p^2} 
              - \frac{{\mathcal{V}^{N S}_{e\,i}}^2}{M_{S_i}} 
              - \frac{{\mathcal{V}^{N N}_{e\,i}}^2}{M_{N_i}}\right).
\label{eq:rr_amplitudes}
\end{eqnarray}
The suitably normalized dimensionless parameters that describe lepton number violation are
\begin{eqnarray}
& &|\eta^{\nu}_{LL}| =  \left|\frac{\sum_i \mathcal{N}_{e\,i}^2 m_i}{m_e} \right| \quad \, , \quad 
               |\eta_{RR}^{\nu}| = \left(\frac{g_{R}}{g_{L}}\right)^4  \left|\left(\frac{\mwl}{\mwr}\right)^4 \frac{\sum_i {\mathcal{V}^{N \nu}_{e\,i}}^2 m_i}{m_e} \right|\, ,  \nonumber \\ 
& &|\eta_{LL}^{S}| =  \left|- m_p \frac{\sum_i {\mathcal{V}^{\nu S}_{e\,i}}^2 }{M_{S_i}} \right| \quad \, , \quad 
               |\eta_{RR}^{S}| = \left(\frac{g_{R}}{g_{L}}\right)^4 \left|- \left(\frac{\mwl}{\mwr}\right)^4 m_p \frac{\sum_i {\mathcal{V}^{N S}_{e\,i}}^2 }{M_{S_i}} \right|\, ,  \nonumber \\ 
& &|\eta_{LL}^{N}| =  \left|- m_p \frac{\sum_i {\mathcal{V}^{\nu N}_{e\,i}}^2 }{M_{N_i}} \right| \quad \, , \quad 
               |\eta_{RR}^{N}| = \left(\frac{g_{R}}{g_{L}}\right)^4  \left|- \left(\frac{\mwl}{\mwr}\right)^4 m_p \frac{\sum_i {\mathcal{V}^{NN}_{e\,i}}^2 }{M_{N_i}} \right|\, ,  \nonumber \\ 
\end{eqnarray}
\subsection{Triplet exchange mechanisms}
Fig \ref{Feyn:all_LR} (iii) is  mediated by $SU(2)_R$ scalar triplet $\Delta_R$ and for this the amplitude is given by  
\begin{equation}
 {\cal A}_{\Delta_R} \simeq G_F^2 \left(\frac{\mwl}{\mwr}\right)^4 \left(\frac{g_{R}}{g_{L}}\right)^4  \sum_i\frac{V^2_{ei}M_i}{m_{\Delta^{--}_R}^2} \propto \frac{L^4}{R^5}\, ,
\label{eq:amp_triplet_R}
\end{equation}
and the dimensionless particle physics parameter is
\begin{equation}
\left|\eta^{\Delta_R}_{RR}\right|  = \left(\frac{\mwl}{\mwr}\right)^4 \left(\frac{g_{R}}{g_{L}}\right)^4 
                \frac{m_p \left|\sum_i V^2_{ei} M_i\right|}{m_{\Delta^{--}_R}^2}\, .
\end{equation}

\subsection{Momentum dependent mechanisms}
In this case the emitted electrons have opposite helicity, and the amplitude is proportional to
\begin{eqnarray}
 {\cal A}_{LR} \simeq & &G_F^2 \left(\left(\frac{\mwl}{\mwr}\right)^2\, \left(\frac{g_{R}}{g_{L}}\right)^2 
                               +\tan\xi 
                               +\left(\frac{\mwl}{\mwr}\right)^2\, \left(\frac{g_{R}}{g_{L}}\right)^2 \tan\xi
                               +\tan^2\!\xi\right)\, 
 \nonumber \\ & & \hspace*{1cm} \times \sum_i\left(\mathcal{V}^{\nu \nu}_{e\,i} {\mathcal{V}^{N \nu}_{e\,i}}^{*}\frac{1}{|p|}
               - {\mathcal{V}^{\nu S}_{e\,i}} {\mathcal{V}^{NS}_{e\,i}}^{*} \frac{|p|}{M^2_{S_i}} 
               - {\mathcal{V}^{\nu N}_{e\,i}} {\mathcal{V}^{NN}_{e\,i}}^{*} \frac{|p|}{M^2_{N_i}}\right);
\label{eq:lr_amplitudes}
\end{eqnarray}
and corresponding dimensionless particle physics parameter involving lepton number violation are 
\begin{eqnarray}
& &|\eta_{\lambda,\nu}| =\left(\frac{g_{R}}{g_{L}}\right)^2 \left|\left(\frac{\mwl}{\mwr}\right)^2 
        \sum_i \mathcal{N}_{e\,i} {\mathcal{V}^{N \nu}_{e\,i}}^{*} \right| \quad \, , \quad \quad \quad 
               |\eta_{\eta,\nu}| = \left|\tan\xi \sum_i \mathcal{N}_{e\,i} {\mathcal{V}^{N \nu}_{e\,i}}^{*}\right|\, ,  \nonumber \\ 
& &|\eta_{\lambda, S}| = \left(\frac{g_{R}}{g_{L}}\right)^2 \left|\left(\frac{\mwl}{\mwr}\right)^2 
        \sum_i {\mathcal{V}^{\nu S}_{e\,i}} {\mathcal{V}^{NS}_{e\,i}}^{*} \frac{|p|^2}{M^2_{S_i}} \right| \quad \, , \quad 
               |\eta_{\eta, S}| = \left|\tan\xi\sum_i {\mathcal{V}^{\nu S}_{e\,i}} {\mathcal{V}^{NS}_{e\,i}}^{*} \frac{|p|^2}{M^2_{S_i}}\right|\, ,  \nonumber \\ 
& &|\eta_{\lambda, N}| = \left(\frac{g_{R}}{g_{L}}\right)^2 \left|\left(\frac{\mwl}{\mwr}\right)^2 
        \sum_i {\mathcal{V}^{\nu N}_{e\,i}} {\mathcal{V}^{NN}_{e\,i}}^{*} \frac{|p|^2}{M^2_{N_i}} \right| \quad \, , \quad 
               |\eta_{\eta,N}| = \left|\tan\xi \sum_i {\mathcal{V}^{\nu N}_{e\,i}} {\mathcal{V}^{NN}_{e\,i}}^{*}\frac{|p|^2}{M^2_{N_i}} \right|\, ,  \nonumber \\ 
\end{eqnarray}

\section{Life time with proper nuclear matrix element and normalized effective mass parameters}
\label{sec:lifetime}
We express the inverse half-life in terms of effective 
mass parameters with proper normalization factors taking into account the nuclear matrix elements \cite{Pantis:1996py,Doi:1985dx,Barry:2013xxa} 
leading to the half-life prediction
\begin{eqnarray}
[T_{1/2}^{0\nu}]^{-1} &=& G_{01}^{0\nu}\bigg\{|{\cal M}_\nu^{0\nu}|^2|\eta^{\rm light}_{LL}|^2 
     + |{\cal M}_N^{0\nu}|^2|\eta^{\rm heavy}_{LL}|^2 \nonumber \\
     &+& |{\cal M}_\nu^{0\nu}|^2|\eta^{\rm light}_{RR}|^2 + |{\cal M}_N^{0\nu}|^2|\eta^{\rm heavy}_{RR}|^2
     +|{\cal M}_\lambda^{0\nu}\eta_\lambda+{\cal M}_\eta^{0\nu}\eta_\eta|^2\bigg\}. 
 \label{eq:halflife_simp}
\end{eqnarray}

$G^{0\nu}_{01}$ is the the phase space factor and matrix elements are 
${\cal M}_k^{0\nu}$ ($k=\nu,N,\lambda,\eta$). Also the dimensionless 
LNV particle physics parameters are 

\begin{align}
|\eta_{\nu}| &= \frac{1}{m_e}\left|\sum_{i=1}^3 {\mathcal{V}^{\nu \nu}_{ei}}^2 m_{\nu_i} \right|\quad \quad  \lesssim {2.66 \times 10^{-7}}, \\[1mm]
|\eta^{\rm heavy}_{LL}| &= m_p \left| - \sum_{i=1}^3 \frac{{\mathcal{V}^{\nu S}_{ei}}^2}{M_{S_i}} 
             -\sum_{i=1}^3 \frac{{\mathcal{V}^{\nu N}_{ei}}^2}{M_{N_i}} \right| \lesssim {2.55 \times 10^{-9}}\, , \\[1mm]
 |\eta^{\rm light}_{RR}| &= \frac{1}{m_e}\left(\frac{\mwl}{\mwr}\right)^4 \left(\frac{g_R}{g_L}\right)^4 
             \left|\sum_{i=1}^3 {\mathcal{V}^{N\nu}_{ei}}^2 m_{\nu_i}\right| \lesssim {2.66 \times 10^{-7}}\, , \label{eq:lightRR}\\[1mm]
 |\eta^{\rm heavy}_{RR}| &= m_p\left(\frac{\mwl}{\mwr}\right)^4 \left(\frac{g_R}{g_L}\right)^4 
            \left| - \sum_{i=1}^3 \frac{{\mathcal{V}^{\nu S}_{ei}}^2}{M_{S_i}} 
             -\sum_{i=1}^3 \frac{{\mathcal{V}^{N N}_{ei}}^2}{M_{N_i}} \right| \lesssim {2.55 \times 10^{-9}}\, ,\\[1mm]
 |\eta_\lambda| &= \left(\frac{\mwl}{\mwr}\right)^2 \left(\frac{g_R}{g_L}\right)^2 
            \left|\sum_{i=1}^3 \mathcal{V}^{\nu N}_{ei} \mathcal{V}^{N N}_{ei} \right| \lesssim {2.18 \times 10^{-7}}\, , \\[1mm]
 |\eta_\eta| &= \tan\xi\left|\sum_{i=1}^3 \mathcal{V}^{\nu N}_{ei} \mathcal{V}^{N N}_{ei} \right| \lesssim {1.13 \times 10^{-9}}\, . 
\end{align}

where, $m_e$ $(m_i)$= mass of electron (light neutrino), and $m_p$ = mass of proton. 
Besides different particle physics parameters, it contains the nuclear matrix elements due to different chiralities of the hadronic 
weak currents such as $\left(\mathcal{M}^{0\nu}_{\nu} \right)$ involving left-left chirality in the standard 
contribution, and $\left(\mathcal{M}^{0\nu}_{\nu} \right)$ involving right-right chirality arising out of heavy neutrino exchange, 
$\left(\mathcal{M}^{0\nu}_{\lambda} \right)$  for the $\lambda$ diagram, and $\left(\mathcal{M}^{0\nu}_{\eta} 
\right)$ for the $\eta$ diagram. It is to be noted here that the current bound on these LNV parameters are derived based on half-life 
limit from the KamLAND-Zen experiment neglecting interference terms.

The numerical values of these nuclear matrix elements as discussed 
in ref.\cite{Pantis:1996py,Doi:1985dx,Barry:2013xxa} are  given in table\,\ref{tab:nucl-matrix}.

\begin{table}[h]
 \centering
 \begin{tabular}{|c|c|c|c|c|c|}
 \hline \hline
{Isotope} & $G^{0\nu}_{01}$  & 
{${\cal M}^{0\nu}_\nu$} &
{${\cal M}^{0\nu}_N$} & 
{${\cal M}^{0\nu}_\lambda$} & 
{${\cal M}^{0\nu}_\eta$}\\
\hline
$^{76}$Ge    & $5.77 \times 10^{-15}$        & 2.58--6.64      & 233--412      & 1.75--3.76   & 235--637 \\ 
$^{136}$Xe   & $3.56 \times 10^{-14}$         & 1.57--3.85      & 164--172      & 1.92--2.49   & 370--419 \\ 
\hline \hline
 \end{tabular}
 \caption{Phase space factors and nuclear matrix elements with their allowed ranges.}
 \label{tab:nucl-matrix}
\end{table}

Using the expression for inverse half-life of $0\nu \beta\beta$ decay process due to only light neutrinos, 
$\left[T_{1/2}^{0\nu}\right]^{-1} = G^{0\nu}_{01}\left|{\cal M}^{0\nu}_\nu \right|^2|\eta_\nu|^2$, we can arrive at a suitable normalization factor for all types of contributions. 
Using the numerical values given in table\,\ref{tab:nucl-matrix}, we rewrite the inverse half-life 
in terms of effective mass parameter as,
$$\left[T_{1/2}^{0\nu}\right]^{-1} = G^{0\nu}_{01} \left| \frac{{\cal M}^{0\nu}_\nu}{m_e} \right|^2\, 
|{\large \bf  m}_{\rm ee}^{\nu}|^2 = 1.57 \times 10^{-25}\, \mbox{yrs}^{-1}\, \mbox{eV}^{-2} |{\large 
\bf  m}_{\rm ee}^{\nu}|^2 = \mathcal{K}_{0\nu}\, |{\large \bf  m}_{\rm ee}^{\nu}|^2 $$
where ${\large \bf  m}_{\rm ee}^{\nu} = \sum^{}_{i}  \left(\mathcal{V}^{\nu \nu}_{e\,i}\right)^2\, m_{\nu_i}$.
Then the analytic expression for all other contributions taking into account the respective 
nuclear matrix elements turns out to be
\begin{eqnarray}
\left[T_{1/2}^{0\nu}\right]^{-1}&=&  \mathcal{K}_{0\nu}\, 
\bigg[ |{\large \bf  m}_{\rm ee}^{\nu}|^2 + |{\large \bf m}_{\rm ee,L}^{S,N}|^2 
  + |{\large \bf  m}_{\rm ee,R}^{S,N}|^2+ |{\large \bf  m}_{\rm ee}^{\lambda}|^2 
  + |{\large \bf  m}_{\rm ee}^{\eta}|^2 \bigg] + \cdots \nonumber \\
&=&\mathcal{K}_{0\nu}\, \bigg[ 
    |{\large \bf  m}_{\rm ee}^{\nu}|^2 + |{\large \bf m}_{\rm ee,L}^{S}+{\large \bf m}_{\rm ee,L}^{N}|^2 
  + |{\large \bf  m}_{\rm ee,R}^{S}+{\large \bf  m}_{\rm ee,R}^{N}|^2 \nonumber \\
  &+& |{\large \bf  m}_{\rm ee}^{\lambda,\nu}+{\large \bf  m}_{\rm ee}^{\lambda,S}+{\large \bf  m}_{\rm ee}^{\lambda,N}|^2 
  + |{\large \bf  m}_{\rm ee}^{\eta,\nu}+{\large \bf  m}_{\rm ee}^{\eta,S}+{\large \bf  m}_{\rm ee}^{\eta,N}|^2 \bigg] + \cdots   
\label{eqn:halflife-eff}
\end{eqnarray}
where the ellipses denote interference terms and all other subdominant contributions. Also the individual effective LNV parameters can be expressed as
\begin{equation}
{\large \bf  m}^{\rm ee}_{\nu} = \sum_{i=1}^3 {\mathcal{V}^{\nu \nu}_{e\,i}}^2\, m_{\nu_i}
\end{equation}
\begin{equation}
{\large \bf  m}_{\rm ee,L}^{S} = \sum_{i=1}^3 {\mathcal{V}^{\nu S}_{e\,i}}^2\, \frac{|p|^2}{M_{S_i}}
\end{equation}
\begin{equation}
{\large \bf  m}_{\rm ee,L}^{N} = \sum_{i=1}^3 {\mathcal{V}^{\nu N}_{e\,i}}^2\, \frac{|p|^2}{M_{N_i}} 
\end{equation}
\begin{equation}
{\large \bf  m}_{\rm ee,R}^{N} = \sum_{i=1}^3 \left(\frac{m_{W_L}}{m_{W_R}}\right)^4 \left(\frac{g_R}{g_L}\right)^4\, 
{\mathcal{V}^{N N}_{e\,i}}^2\, \frac{|p|^2}{M_{N_i}} 
\end{equation}
\begin{equation}
{\large \bf  m}_{\rm ee}^{\lambda,\nu} =10^{-2}\, \left(\frac{\mwl}{\mwr}\right)^2\, \left(\frac{g_R}{g_L}\right)^2\, |p|\,
           \sum_{i=1}^3 \left[ U_{\rm PMNS} \frac{M_D}{M_N} \cdots \right]_{ee}
\end{equation}
\begin{equation}
{\large \bf  m}_{\rm ee}^{\lambda,S} =10^{-2}\, \left(\frac{m_{W_L}}{m_{W_R}}\right)^2 \left(\frac{g_R}{g_L}\right)^2\,
\sum_{i=1}^3 \mathcal{V}^{\nu S}_{e\,i} \mathcal{V}^{N S}\, \frac{|p|^3}{M^2_{S_i}}
\end{equation}
\begin{equation}
{\large \bf  m}_{\rm ee}^{\lambda,N} =10^{-2}\, \left(\frac{m_{W_L}}{m_{W_R}}\right)^2 \left(\frac{g_R}{g_L}\right)^2\, 
\sum_{i=1}^3 \mathcal{V}^{\nu N}_{e\,i} \mathcal{V}^{N N}\, \frac{|p|^3}{M^2_{N_i}}
\end{equation}
\begin{equation}
{\large \bf  m}_{\rm ee}^{\eta} = \tan \zeta_{LR}\, |p|\,
          \sum_{i=1}^3 \left[ U_{\rm PMNS} \frac{M_D}{M_N} \cdots \right]_{ee}
\end{equation}
where $\langle p \rangle^2 = - m_e\,m_p\, \mathcal{M}^{0\nu}_{N}/\mathcal{M}^{0\nu}_{\nu} \simeq |\mbox{200\, MeV}|^2$. 
It is to be noted that the suppression factor $10^{-2}$ arises in the $\lambda$diagram because of normalization w.r.t. the standard mechanism.

\section{The role of Pati-Salam symmetry}
\label{sec:unification}
We know that both the gauge couplings for $SU(2)_L$ and $SU(2)_R$ are exactly equal at a scale when either Pati-Salam symmetry with discrete left-right symmetry $SU(2)_L \times SU(2)_R \times SU(4)_C \times D \equiv \mathcal{G}_{224D}$ or manifest left-right symmetry $SU(2)_L \times SU(2)_R \times U(1)_{B-L} \times SU(3)_C \times D \equiv \mathcal{G}_{2213D}$ appears as an intermediate symmetry breaking step. This equality is sustained as long as D-parity 
remains unbroken. Once the spontaneous breaking of D-parity occurs, it immediately results in  
$g_{L} \neq g_{R}$ and the ratio $\frac{g_L}{g_R}$ deviates from unity depending upon the 
breaking scale. In the considered model we have found this ratio $\frac{g_L}{g_R}$ 
to be $\simeq 1.5$ which will be supportive in predicting new non-standard contributions to neutrinoless double 
beta decay. The deviation of this ratio from unity is enhanced by the occurrence of Pati-Salam symmetry as an 
intermediate scale and thus justifies its importance in explaining $0\nu\beta\beta$, LFV decays 
as well as collider processes within the framework of $SO(10)$ GUT based models. The importance of Pati-Salam symmetry as the highest 
intermediate step in a SO(10) symmetry breaking chain has already been discussed in ref \cite{Parida:1996td}. For quantifying these points, 
we consider the following non-SUSY $SO(10)$ chain, as an example.
\begin{eqnarray}
&&\quad \quad SO(10) \mathop{\longrightarrow}^{M_U}_{} \mathcal{G}_{224D} \,
       \mathop{\longrightarrow}^{M_P}_{} \mathcal{G}_{224} \,
       \mathop{\longrightarrow}^{M_C}_{} \mathcal{G}_{2213} \,
       \mathop{\longrightarrow}^{M_R}_{} \mathcal{G}_{SM}\,
       \mathop{\longrightarrow}^{M_{Z}}_{}\mathcal{G}_{13}\, . \nonumber 
\end{eqnarray}
It was found that the $G_{224}$-singlets contained in $\{54\}_H$ and $\{210\}_H$ of $SO(10)$ are D-parity 
even and odd, respectively. Moreover the neutral components of the $G_{224}$ multiplet 
$\{1, 1, 15\}$ contained in $\{210\}_H$ and $\{45\}_H$ of $SO(10)$ were also found to be D-parity even and odd, 
respectively. Here in the first step, VEV is assigned to the $\langle(1, 1, 1)\rangle\subset\{54\}_H$ 
which has even D-Parity to ensure the survival of LR symmetric Pati-Salam group while in the second 
step D-parity is broken by assigning $\langle(1, 1, 1) \rangle \subset\{210\}_H$ to obtain asymmetric 
$G_{224}$ with $g_{L}\neq g_{R}$. Then the spontaneous breaking of $G_{224}\rightarrow G_{2213}$ is achieved 
by the VEV $\langle(1, 1, 15)^0_H\rangle\subset\{210\}_H$. The breaking of $SU(2)_R\times U(1)_{B-L}\rightarrow 
U(1)_Y$ is achieved by $\langle\Delta^0_R\rangle \subset\{126\}_H$ while the VEV $\langle\chi_R^0\rangle 
\subset \{16\}_H$ provides the $N$-$S$ mixing. Finally, as usual, the breaking of SM to low energy theory 
$U(1)_{\text{em}}\times  SU(3)_C$ is carried out by the SM bidoublet $\Phi\subset \{10\}_H$.

\subsection{Gauge coupling unification}
We consider three different cases for gauge coupling unification as follows and we also show the Higgs spectrum used in different 
ranges of mass scales under respective gauge symmetries.\\
{\bf Case - I :} Symmetric LR model ($g_L = g_R$) 
\begin{eqnarray}
&&\quad \quad SO(10)\mathop{\longrightarrow}^{M_U}_{} \mathcal{G}_{2213D} \,
       \mathop{\longrightarrow}^{M_{R}}_{}\mathcal{G}_{SM}\,
       \mathop{\longrightarrow}^{M_{Z}}_{}\mathcal{G}_{13}\, . \nonumber 
\end{eqnarray}
\begin{eqnarray}
& &\hspace*{-1.0cm} {\bf \mbox{(i)}\, \mu=M_Z - M_{R} }:  \mathcal{G}={\rm SM} = \mathcal{G}_{213}, 
             \hspace*{0.2cm} \phi (2,1/2,1)\, ; \nonumber \\
& &\hspace*{-1.0cm} {\bf \mbox{(ii)}\, \mu=M_{R} - M_U}: \mathcal{G}= \mathcal{G}_{2213}, 
             \hspace*{0.2cm} \Phi_1 (2,2,0,1) \oplus \Phi_2 (2,2,0,1) \oplus \chi_L (2,1,-1,1) \oplus \nonumber \\ 
& &          \hspace*{4.9cm} \chi_R (1,2,-1,1) \oplus \Delta_L (3,1,2,1) \oplus \Delta_R (1,3,2,1) \oplus
\nonumber \\
& &          \hspace*{4.9cm} 4\delta^{+} (1,1,2,1) \oplus \eta (1,1,-2/3,3) \oplus \xi (1,1,4/3,6)   
\end{eqnarray}
{\bf Case - II :} Asymmetric LR model ($g_L \neq g_R$)
\begin{eqnarray}
&&\quad \quad SO(10) \mathop{\longrightarrow}^{M_U}_{} \mathcal{G}_{2213D} \,
       \mathop{\longrightarrow}^{M_C}_{} \mathcal{G}_{2213} \,
       \mathop{\longrightarrow}^{M_{R}}_{}\mathcal{G}_{SM}\,
       \mathop{\longrightarrow}^{M_{Z}}_{}\mathcal{G}_{13}\, . \nonumber 
\end{eqnarray}
\begin{eqnarray}
& &\hspace*{-1.0cm} {\bf \mbox{(i)}\, \mu=M_Z - M_{R} }:  \mathcal{G}={\rm SM} = \mathcal{G}_{213}, 
             \hspace*{0.2cm} \phi (2,1/2,1)\, ; \nonumber \\
& &\hspace*{-1.0cm} {\bf \mbox{(ii)}\, \mu=M_{R} - M_U}: \mathcal{G}= \mathcal{G}_{2213}, 
             \hspace*{0.2cm} \Phi_1 (2,2,0,1) \oplus \Phi_2 (2,2,0,1) \oplus \chi_R (1,2,-1,1) \oplus \nonumber \\ 
& &          \hspace*{4.9cm} \Delta_R (1,3,2,1) \oplus 6\delta^{+} (1,1,2,1)       
\end{eqnarray}
Now, we have introduced the Pati-Salam symmetry in the $SO(10)$ symmetry breaking chain. We have divided case-III into IIIA and IIIB where IIIA stands for the case where $SO(10)$ and $D$-parity break at same scale ($M_U \approx M_P$) and in IIIB, we have presented the $D-$parity breaking at some lower scale than $M_U$.\\  
{\bf Case - IIIA :} 
\begin{eqnarray}
&&\quad \quad SO(10) \mathop{\longrightarrow}^{M_U \approx M_P}_{} \mathcal{G}_{224} \,
       \mathop{\longrightarrow}^{M_C}_{} \mathcal{G}_{2213} \,
       \mathop{\longrightarrow}^{M_R}_{} \mathcal{G}_{SM}\,
       \mathop{\longrightarrow}^{M_{Z}}_{}\mathcal{G}_{13}\, . \nonumber 
\end{eqnarray}
\begin{eqnarray}
& &\hspace*{0.5cm} {\bf \mbox{(i)}\, \mu=M_Z - M_{R} }:  \mathcal{G}={\rm SM} = \mathcal{G}_{213}, 
             \hspace*{0.2cm} \phi (2,1/2,1)\, ; \nonumber \\
& &\hspace*{0.5cm} {\bf \mbox{(ii)}\, \mu=M_R - M_C}: \mathcal{G}= \mathcal{G}_{2213},
             \hspace*{0.2cm} \Phi_1 (2,2,0,1) \oplus \Phi_2 (2,2,0,1) \oplus \chi_L (2,1,-1,1)\oplus
\nonumber \\ 
& &          \hspace*{6.5cm} \chi_R (1,2,-1,1)\oplus \Delta_L (3,1,-2,1)\oplus \Delta_R (1,3,-2,1)\, ; \nonumber \\ 
& &\hspace*{0.5cm} {\bf \mbox{(iii)}\, \mu=M_C - M_U}: \mathcal{G}= \mathcal{G}_{224},
             \hspace*{0.3cm} \Phi_1 (2,2,1) \oplus \Phi_2 (2,2,1) \oplus \chi_R (1,2, {\bar 4}) \oplus \Delta_R (1,3,{\bar {10}}) \oplus
             \nonumber \\ 
& &          \hspace*{6.5cm} \Omega_R (1,3,15) \oplus \Sigma(1,1,15) \oplus \zeta(2,2,15)\oplus \eta (2,2,1) \oplus
\nonumber \\ 
& &          \hspace*{6.5cm} \rho (2,1,4)\,.
\end{eqnarray}
{\bf Case - IIIB :} 
\begin{eqnarray}
&&\quad \quad SO(10) \mathop{\longrightarrow}^{M_U}_{} \mathcal{G}_{224D} \,
       \mathop{\longrightarrow}^{M_P}_{} \mathcal{G}_{224} \,
       \mathop{\longrightarrow}^{M_C}_{} \mathcal{G}_{2213} \,
       \mathop{\longrightarrow}^{M_R}_{} \mathcal{G}_{SM}\,
       \mathop{\longrightarrow}^{M_{Z}}_{}\mathcal{G}_{13}\, . \nonumber 
\end{eqnarray}
\begin{eqnarray}
& &\hspace*{0.5cm} {\bf \mbox{(i)}\, \mu=M_Z - M_{R} }:  \mathcal{G}={\rm SM} = \mathcal{G}_{213}, 
             \hspace*{0.2cm} \phi (2,1/2,1)\, ; \nonumber \\
& &\hspace*{0.5cm} {\bf \mbox{(ii)}\, \mu=M_R - M_C}: \mathcal{G}= \mathcal{G}_{2213},
             \hspace*{0.2cm} \Phi_1 (2,2,0,1) \oplus \Phi_2 (2,2,0,1) \oplus \chi_L (2,1,-1,1)\oplus
\nonumber \\ 
& &          \hspace*{6.5cm} \chi_R (1,2,-1,1)\oplus \Delta_L (3,1,-2,1)\oplus \Delta_R (1,3,-2,1)\, ; \nonumber \\ 
& &\hspace*{0.5cm} {\bf \mbox{(iii)}\, \mu=M_C - M_P}: \mathcal{G}= \mathcal{G}_{224},
             \hspace*{0.3cm} \Phi_1 (2,2,1) \oplus \Phi_2 (2,2,1) \oplus \chi_R (1,2, {\bar 4}) \oplus \Delta_R (1,3,{\bar {10}}) \oplus
             \nonumber \\ 
& &          \hspace*{6.5cm} \Omega_R (1,3,15) \oplus \Sigma(1,1,15) \oplus \xi(2,2,15)\, ; \nonumber \\ 
& &\hspace*{0.5cm} {\bf \mbox{(iv)}\, \mu=M_P - M_U}: \mathcal{G}= \mathcal{G}_{224D},
             \hspace*{0.2cm} \Phi_1 (2,2,1) \oplus \Phi_2 (2,2,1) \oplus \chi_L (2,1,4) \oplus \chi_R (1,2, {\bar 4}) \oplus \nonumber \\ 
& &          \hspace*{6.5cm} \Delta_L (3,1,10) \oplus \Delta_R (1,3,{\bar {10}}) \oplus \Omega_L(3,1,15)\oplus \Omega_R (1,3,15)\oplus \nonumber \\ 
& &          \hspace*{6.5cm} \Sigma(1,1,15) \oplus \xi(2,2,15)\ \oplus \sigma(1,1,1)\,.
\end{eqnarray}
\begin{figure}[h]
\centering
\includegraphics[scale=0.37]{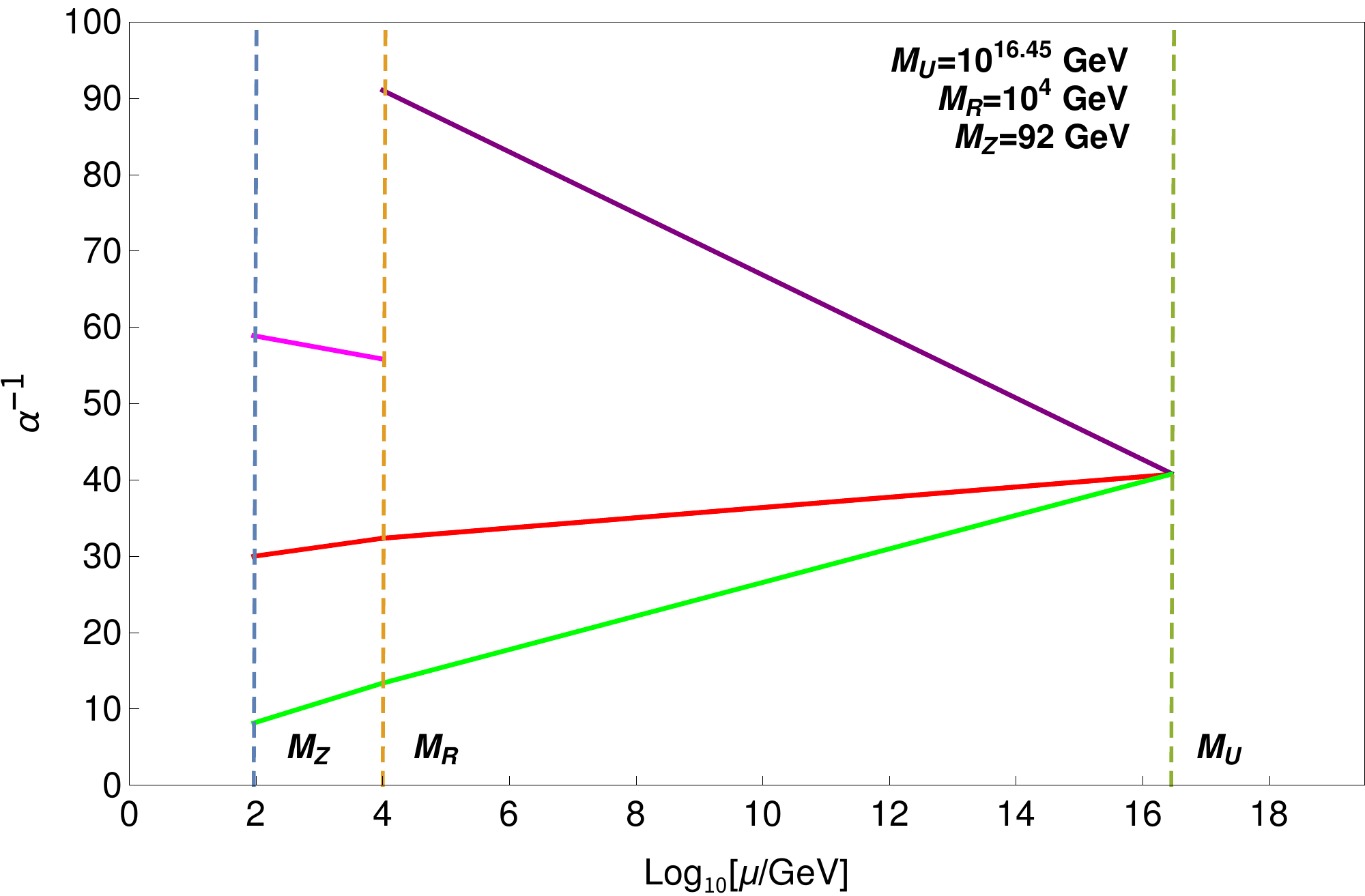}\qquad
\includegraphics[scale=0.37]{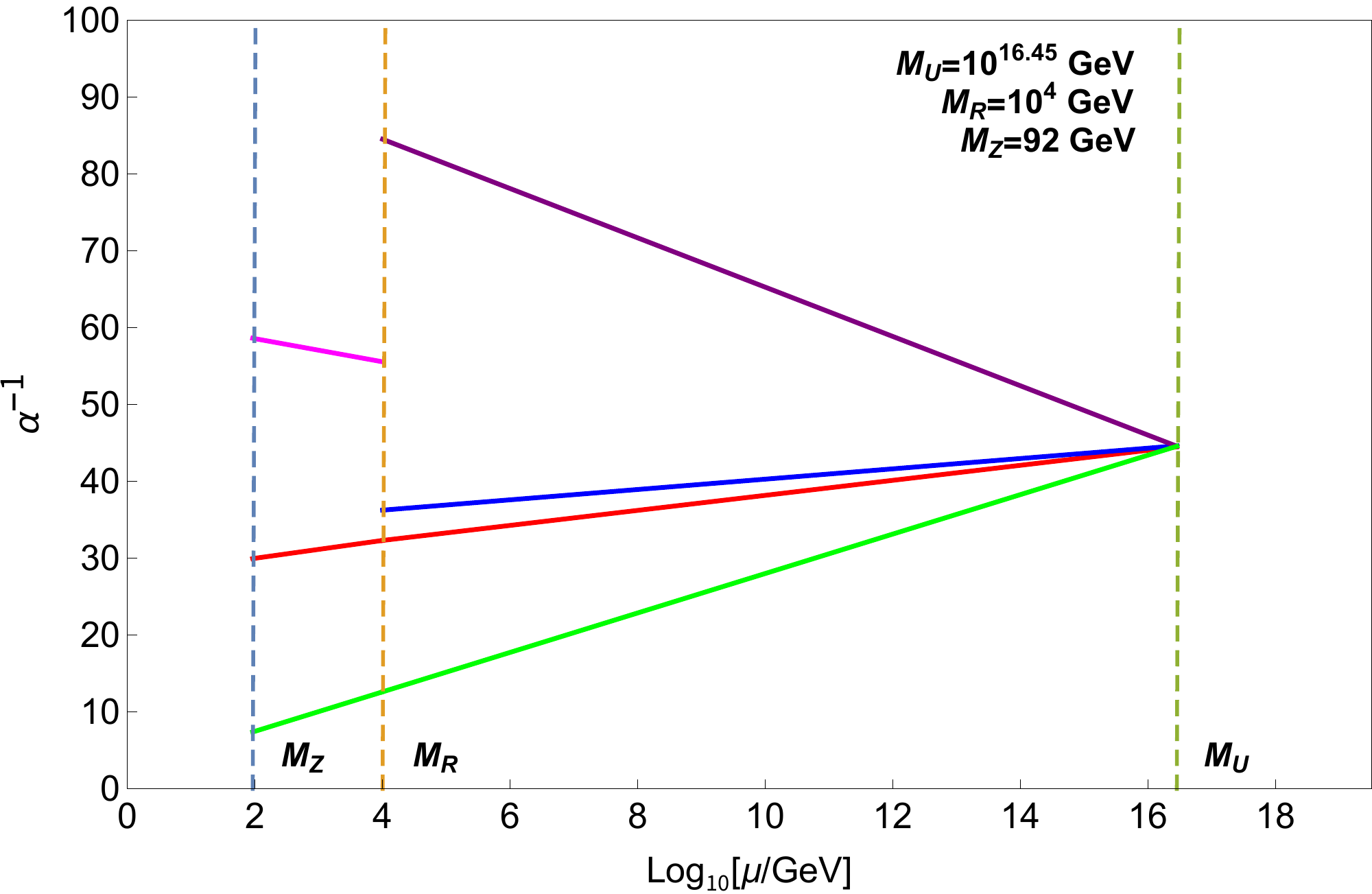}
\caption{The plot in the left panel shows gauge coupling unification at the scale $M_U=10^{16.45}~ \text{GeV}$ for the symmetric LRSM case, i.e. $g_L=g_R$ (case-I). 
The plot in the right panel shows gauge coupling unification at the scale $M_U=10^{16.45}~ \text{GeV}$ for the asymmetric LRSM case, 
i.e. $g_L\neq g_R$ without Pati-Salam symmetry (case-II).}
\label{unif_1,2}
\end{figure}
\begin{figure}[h]
\centering
\includegraphics[scale=0.37]{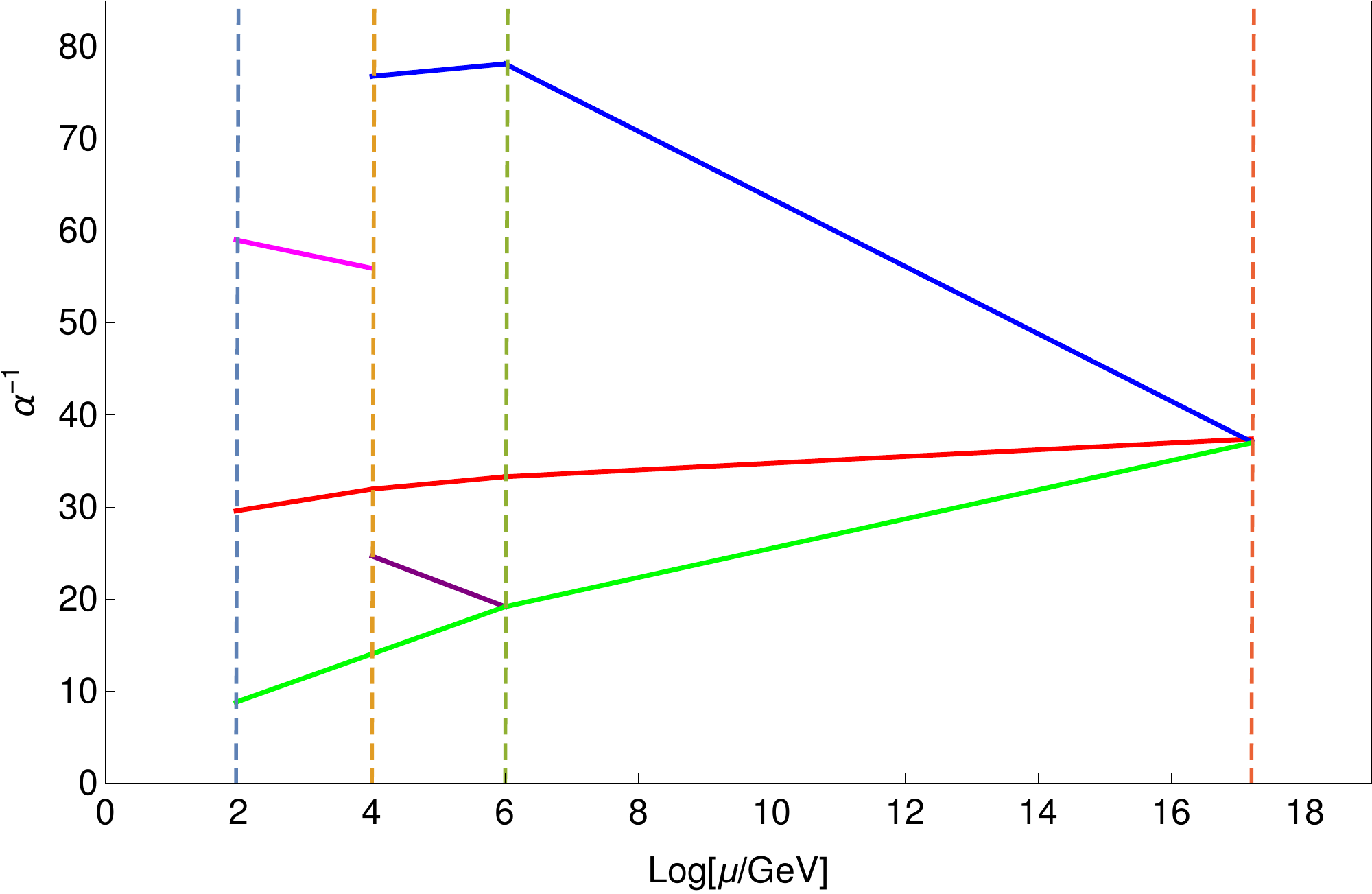}\qquad
\includegraphics[scale=0.37]{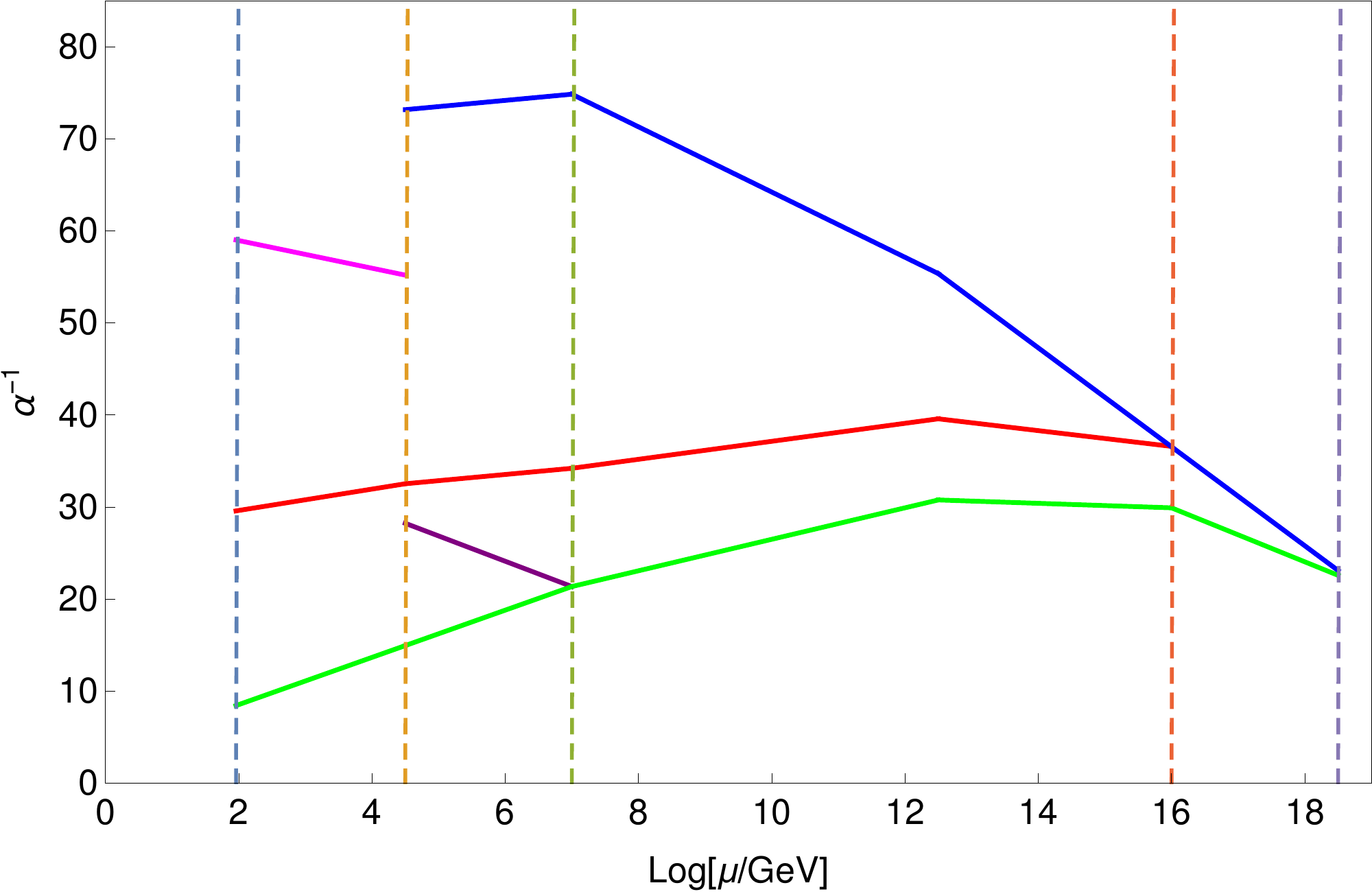}
\caption{The left panel plot shows the gauge coupling unification as well as D-parity breaking at $M_U \sim M_P = 10^{17.2}$~GeV (Case-IIIA). 
In the right panel, we present the unification at about $M_U = 10^{18.5}$~ GeV after introducing Pati-Salam symmetry as the highest intermediate 
symmetry breaking scale with D-parity breaking at around $M_P = 10^{16}$~GeV (Case-IIIB). For both the cases, $g_L \neq g_R$.}
\label{unif_3,4}
\end{figure}
The gauge coupling unification plots for the above four cases are shown in Fig.\,\ref{unif_1,2} and Fig.\,\ref{unif_3,4} respectively. 
In the unification plots the different colored lines stand for running of various gauge groups. The red, blue, pink, magenta and green 
lines are for $SU(2)_L$, $SU(2)_R$, $U(1)_Y$, $U(1)_{B-L}$, $SU(3)_C$ gauge groups respectively. 
For case-IIIB we have added an extra particle $\xi(2,2,15)$ which helps us to unify the $SU(2)_L$ and $SU(2)_R$ gauge couplings at around $10^{16}$ GeV (i.e, the D-parity breaking scale of $\mathcal{G}_{224D}$), also this extension of the model gives us the advantage to acquire fermion mass fitting at GUT scale.

\pagebreak

%%%%%%%%%%%%%%%%%%%%%%%%%%%%%%%%%%%%%%%%%%%%%%%%%%%%%%%%%%%%%%%%%

\bibliographystyle{utcaps_mod}
\bibliography{LNV}
\end{document}